\documentclass[10pt,twocolumn,oneside,final]{IEEEtran}

\usepackage{cite}
\usepackage{graphicx}
\usepackage{amsmath}
\usepackage{times}
\usepackage{latexsym}
\usepackage{bm}
\usepackage{amssymb}
\usepackage[center]{caption2}
\usepackage{array}
\usepackage{setspace}
\usepackage{fancyhdr}
\usepackage{cite,graphicx,amsmath,amssymb}
\usepackage{citesort}

\ifCLASSINFOpdf

\else

\fi

\newtheorem{theorem}{Theorem}
\newtheorem{lemma}{Lemma}

\newtheorem{proposition}{Proposition}
\newtheorem{alg}{Algorithm}

\newcommand{\tr}{\text{tr}}

\usepackage[dvipsnames,usenames]{color}
\usepackage[usenames,dvipsnames]{color}

\makeatother

\begin{document}
%
%

\title{Transmit Designs for the MIMO Broadcast Channel with Statistical CSI}

\author{Yongpeng Wu, \IEEEmembership{Member, IEEE}, Shi Jin, \IEEEmembership{Member, IEEE},
Xiqi Gao, \IEEEmembership{Senior Member, IEEE}, \\ Matthew R. McKay,
\IEEEmembership{Senior Member, IEEE},  and Chengshan Xiao, \IEEEmembership{Fellow, IEEE}.

\thanks{Copyright (c) 2014 IEEE. Personal use of this material is permitted. However, permission to use this material
for any other purposes must be obtained from the IEEE by sending a request to pubs-permissions@ieee.org.}

\thanks{Part of this paper was presented at IEEE International Conference on Communication (ICC'11), Kyoto, Japan, Jun. 2011. }

\thanks{The work of Y. Wu, S. Jin, and X. Gao was supported in part by
 National Natural Science Foundation of China under Grants 61320106003 and 61222102,
the China High-Tech 863 Plan under Grant 2012AA01A506, National Science and Technology Major Project of China under Grants 2013ZX03003004
and 2014ZX03003006-003, the Program for Jiangsu Innovation Team, and the Supporting Program for
New Century Excellent Talents in University.
The work of Y. Wu was also supported
by the Alexander von Humboldt Foundation.
The work of M. R. McKay was supported by the Hong Kong Research
Grants Council under grant number 616713.
The work of C. Xiao was supported in part by National Science
Foundation under Grants CCF-0915846 and ECCS-1231848.
Part of this
work was carried out while Y. Wu was a visiting scholar
at Missouri University of Science and Technology.
}

\thanks{Y. Wu was
with the National Mobile Communication Research Laboratory,
Southeast University, Nanjing, 210096, P. R. China (email: ypwu@seu.edu.cn).
Y. Wu is now with Institute for Digital Communications, Universit$\ddot{a}$t Erlangen-N$\ddot{u}$rnberg,
Cauerstrasse 7, D-91058 Erlangen, Germany (email: yongpeng.wu@lnt.de).  }

\thanks{S. Jin and X. Gao are with the National Mobile Communication Research Laboratory,
Southeast University, Nanjing, 210096, P. R. China (email: jinshi@seu.edu.cn; xqgao@seu.edu.cn;). }
\thanks{M. R. McKay is with Department of Electrical and Computer Engineering,
Hong Kong University of Science and Technology, Clear Water Bay, Kowloon, Hong Kong, (email: eemckay@ust.hk).}
\thanks{C. Xiao is with Department of Electrical and Computer Engineering,
Missouri University of Science and Technology, Rolla, MO 65409, USA, (email: xiaoc@mst.edu).}
}





\maketitle

\begin{abstract}
We investigate the multiple-input multiple-output broadcast channel
with statistical channel state information  available at the transmitter.
The so-called \emph{linear assignment} operation is employed, and
necessary conditions are derived for the optimal transmit design under general fading conditions.
Based on this, we introduce an iterative algorithm to maximize
the  linear assignment weighted sum-rate by applying a gradient descent method.
 To reduce complexity, we
derive an upper bound of the linear assignment achievable rate of each receiver, from which
a simplified closed-form expression for a near-optimal
linear assignment matrix is derived. This reveals an
interesting  construction analogous to that of  dirty-paper coding.
In light of this, a low complexity transmission scheme is provided.
Numerical examples
illustrate the significant performance of the proposed low complexity scheme.

\end{abstract}



\section{Introduction}\label{sec:introduction}
The multiple-input multiple-output (MIMO) broadcast channel (BC) with
Gaussian noise has attracted tremendous research interest in recent
years. Dirty paper coding (DPC) has been proved to achieve the
capacity region of this channel \cite{Weingarten}, whereas various
linear precoding techniques have also been developed to reduce
complexity (see, e.g., \cite{Spencer,Yoo,Sadek,Chae,Stankovic,Wu2012TWC}).
Much current work dealing with the MIMO BC model assume
that  {instantaneous} channel state information (CSI) is
available at both the transmitter and receivers, in order to fully
capitalize on the spatial multiplexing advantages of the MIMO
transmission. Whilst this assumption may be plausible for fixed or
low mobility applications for which the channel realizations change
slowly enough to be monitored at the transmitter (e.g., via a
feedback link or by exploiting channel reciprocity), for other
applications it becomes less feasible.  In particular, as mobility
increases, the channel fluctuations begin to vary more rapidly, and
tracking these gains accurately at the transmitter becomes
problematic.

For mobile applications, an alternative approach is to exploit
 statistical CSI at the transmitter
\cite{Tulino,Piantanida,Piantanida_2,Chalise,Xu2009TVT,Bjornson_2,Raghavan,Wu2014}; a technique
which has drawn much attention in  MIMO system design recently \cite{Gao}. Compared with instantaneous CSI,
the statistical parameters typically vary over a much
longer time window, and therefore can be monitored  more easily
at the transmitter. With statistical CSI at the transmitter, an important problem is to understand
the information-theoretic limits of the MIMO BC model.
In the special case of ``more capable" channels,  where the power of signals to each receiver can be ordered, the ergodic capacity region
was analyzed in \cite{Gammal}, developing upon earlier work \cite{Cover,Costa}. For the fading single-input single-output (SISO) BC model,
an achievable inner bound of the ergodic capacity region was proposed in \cite{Tuninetti}. This work was extended to the fading multiple-input single-output (MISO)
BC  model in \cite{Jafar}, where the distributions of the fading coefficients were assumed isotropic and
the ergodic capacity region was proved to collapse to that of the fading SISO  BC model. Zhang \textit{et al.}  \cite{Zhang}
examined an outage achievable rate region for the fading single-input multiple-output (SIMO) and MISO  BC models. Very recently,
simple linear precoding designs for some special MISO BC models were proposed in \cite{Ghosh,JWang2012TSP}.

Despite significant advances as described above, for the
fading MIMO BC model, the capacity region remains unknown.
To simplify the problem, a so-called linear assignment operation was proposed in \cite{Bennatan}.
Also,
a linear assignment capacity was defined in \cite{Bennatan} which,  to the best of our knowledge,
is the most systematic result revealing the information-theoretic limits
of the fading MIMO BC model so far.  However,  comprehensive and explicit transmit designs based on this
linear assignment operation were not given in \cite{Bennatan} and are still missing in general.

{
In this paper, starting with the definition of the linear assignment
capacity in  \cite{Bennatan}, we consider the transmit design problem aimed at optimizing the
 linear assignment weighted sum-rate (LAWSR)  of the
fading MIMO BC model with statistical CSI at the transmitter.
Based on an exact expression of the LAWSR, we reveal two key elements that need
to be properly designed: 1) The linear assignment matrices; 2) The precoding matrices of the receivers.

We make the following key contributions:

\begin{enumerate}

\item[1)] We establish
necessary conditions for the optimal linear assignment matrices and precoding matrices.
  Accordingly, the joint design of these can be formulated as a
multidimensional optimization problem, which is solved by an alternating
optimization method with a gradient descent update.

\item[2)] For the linear assignment matrix
of each receiver,  we provide a heuristic design with reduced computational complexity.
This is done by rewriting the expression for the
linear assignment achievable rate (LAAR) of each receiver as a difference of two terms, and applying Jensen's inequality to each.
This operation results in  similar ``bounding errors" for each of the two terms, but
we prove it still leads to
a strict upper bound of LAAR, which turns out to be fairly tight.

\item[3)] The derived upper bound motivates the establishment of a near-optimal construction
for the linear assignment matrix to maximize the LAAR.
Based on this construction, a simplified closed-form
expression for the linear assignment matrix via second-order statistics of the CSI
is obtained. This expression resembles the DPC structure, where
the interference signal power has no impact on determining the linear assignment matrix.
Moreover, for the design based on the derived upper bound and  the simplified linear assignment matrix,
it is shown that an interference elimination effect analogous to that of DPC
transmission  with instantaneous CSI exists.  In light of this, a low complexity
algorithm without numerical averaging is proposed to design the precoding matrices.

\item[4)] We reveal that if all the
channels of the receivers are independent and identically distributed (i.i.d.) fading,
the time division multiple access (TDMA) transmission is optimal. Also, we reveal that
if the transmitter has only one antenna, the opportunistic scheduling transmission
based on the statistical CSI is optimal\footnote{For the single antenna i.i.d. case, the TDMA transmission
and the opportunistic scheduling transmission are equivalent.}.

\end{enumerate}

Numerical results are presented to examine the proposed transmission designs.
These indicate that the proposed designs perform close
to the no-interference upper bound of the MIMO BC with statistical CSI \cite{Bennatan}, and achieve significant performance gains compared
to the TDMA transmission in various scenarios. }

The reminder of the paper is organized as follows. Section II describes the fading MIMO BC model under consideration. In Section III, we establish necessary conditions of the optimal
linear assignment matrices and  precoding matrices for all the receivers, and propose an iterative
algorithm to search for the optimal solution.  In Section IV, we derive an upper bound of the LAAR of each receiver, based on which we investigate low complexity transmit
designs.  Numerical results are provided in Section V and the main results are summarized in Section VI. Main mathematical proofs have been placed in the Appendices.

The following notation is adopted throughout the paper: Vectors are represented as columns and are denoted in lower case bold-face,
and matrices are represented in upper case bold-face.   The superscripts $(\cdot)^{T}$, $(\cdot)^{*}$, and $(\cdot)^{H}$ stand for the matrix transpose, conjugate and conjugate-transpose operations, respectively. We use $\rm{det}(\cdot)$ and $\rm{tr}(\cdot)$ to denote the matrix determinant and trace operations respectively, and ${\bf{A}}^{-1}$ denote the inverse of matrix $\bf{A}$. $\circ$ denotes the Hadamard product of two matrices.
$\mathbf{A} \succeq \mathbf{0}$ means that $\mathbf{A}$ is  Hermitian positive semi-definite, and $\mathbf{A} \succ \mathbf{0}$ means that $\mathbf{A}$ is Hermitian positive definite.
$\left\| {\mathbf{X}} \right\|_F$ denotes the Frobenius norm of matrix $\mathbf{X}$.  The $M \times M$ identity matrix is denoted by ${\bf{I}}_M$, and the all-zero matrix is denoted by $\bf{0}$. The complex number field is represented by $\mathbb{C}$, and $E\left[\cdot\right]$ evaluates the expectation of all the random variables within the bracket.

\section{System Model}\label{sec:system_model}
Consider a fading MIMO BC scenario, where the
transmitter has $N_t$ antennas, while each of the $L$ receivers has
$N_r$ antennas.  The received vector for the $l$-th receiver can be written as
\begin{equation}\label{model}
{\bf{y}}_l  = {\bf{H}}_l {\bf{x}}_T
 + {\bf{z}}_l
\end{equation}
where
${\bf{x}}_T \in {\mathbb{C}}^{{N_t  \times 1} }$ is the
transmitted vector designed to meet the power constraint
\begin{equation}\label{power_constraint}
E\left[ {{\rm{tr}}\left( {{\bf{x}}_T{\bf{x}}_T^H } \right)} \right] \le P .
\end{equation}
${\bf{z}}_l \in {\mathbb{C}}^{ {N_r \times 1}}$ contains
circularly symmetric complex Gaussian noise with zero-mean and
covariance  $E\left[ {{\bf{z}}_l {\bf{z}}_l ^H } \right] $ $= N_0 {\bf{I}}_{N_r}$, and ${\bf{H}}_l
 \in {\mathbb{C}}^{{N_r  \times N_t} } $ is a random channel matrix. The
channels are assumed independent across $l$, i.e., for different receivers.
Moreover,  the $l$-th receiver is assumed to
perfectly estimate its own channel matrix, ${\bf{H}}_l$,
whilst the transmitter only knows statistical CSI\footnote{It is noted that
throughout this paper, we make no assumption on the distribution of $\mathbf{H}_l$,
beyond it having finite energy.} for each receiver.

Here, the transmitted signal ${\bf{x}}_T$ in (\ref{model}) is constructed as ${\bf{x}}_T = \sum\nolimits_{l = 1}^L {{\bf{x}}_l }$,
where ${\bf{x}}_l$ contains the transmitted signal destined for the $l$-th receiver with the
covariance matrix ${\bf{\Sigma }}_l = E\left[ {{\bf{x}}_l{\bf{x}}_l^H } \right]$. Similar to \cite{Bennatan}, we assume that ${\bf{\Sigma}} _l$ is nonsingular.
{ Moreover, we follow \cite{Bennatan} to assume that the transmitter  generates the transmitted data for each user in ascending order.} Thus,
when  designing the transmitted signal for the $l$-th receiver, the transmitter possesses the full non-causal knowledge of the transmitted codewords
for the receivers $1,2,\cdots,l-1$. To this end, we reexpress (\ref{model}) as
\begin{equation}\label{model_2}
{\mathbf{y}}_l = {\mathbf{H}}_l \left( {{\mathbf{x}}_l  +  \mathbf{s}_l } \right) + {\mathbf{H}}_l \sum\limits_{t = l + 1}^L {{\mathbf{x}}_t }  + {\mathbf{z}}_l
\end{equation}
where $\mathbf{s}_l = \sum\limits_{t = 1}^{l - 1} {{\mathbf{x}}_t }$.  In the absence of the instantaneous CSI at the transmitter, the pair $(\mathbf{y}_l,\mathbf{H}_l)$ constitutes the channel output.
The channel transition probability ${\rm{Pr}}[\mathbf{y}_l, \mathbf{H}_l|\mathbf{x}_l,\mathbf{s}_l]$ of the fading MIMO BC model in (\ref{model_2}) is a function of the transmitted signal ${\mathbf{x}}_l$, the channel matrix
$\mathbf{H}_l$, and the state of the non-causally known interference $\mathbf{s}_l$, which is an instance of the general class of
side-information channels \cite{Shannon,Gel}.  Hence the capacity for the $l$-th receiver in model (\ref{model_2}) is defined as
\begin{equation}\label{fading_capacity}
C_l = \mathop {\sup }\limits_{\Pr [{\mathbf{u}}_l\left| {\mathbf{s}_l} \right.], \, f_l\left( . \right)} \left\{ {I\left( {{\mathbf{u}}_l;{\mathbf{y}_l},{\mathbf{H}_l}} \right) - I\left( {{\mathbf{u}_l};{\mathbf{s}_l}} \right)} \right\}
\end{equation}
where ${\mathbf{u}_l}$ is an auxiliary random vector with the conditional distribution $\Pr [{\mathbf{u}_l}\left| {\mathbf{s}_l} \right.]$ and $f_l\left( \cdot \right)$ is a deterministic
function which constructs the transmitted signal as ${\mathbf{x}_l} = f_l({\mathbf{u}_l,\mathbf{s}_l})$ and satisfies $E\left[ { f_l({\mathbf{u}_l,\mathbf{s}_l})} { f_l({\mathbf{u}_l,\mathbf{s}_l})}^H \right] = {{\bf{\Sigma }}_l}$.
We note that for any particular choice of $\Pr [{\mathbf{u}_l}\left| {\mathbf{s}_l} \right.]$ and $f_l\left( \cdot \right)$,
expression (\ref{fading_capacity}) becomes an achievable transmission rate.

The maximum achievable rate of the fading MIMO BC model in (\ref{fading_capacity}) remains an open problem.
However, by generating $\mathbf{x}_l = f_l({\mathbf{u}_l,\mathbf{s}_l}) = {\mathbf{u}_l} - \mathbf{F}_l \mathbf{s}_l$,
where ${\mathbf{F}}_l$ represents the $N_t \times N_t$ constant
 \emph{linear assignment} matrix, we have an achievable transmission rate
  \begin{equation}\label{fading_rate}
R_{{\rm{LA}}, \, l} = \mathop {\sup }\limits_{\mathbf{F}_l} \left\{ R_{l}\right\}
\end{equation}
where
\begin{equation}\label{fading_rate_la}
\begin{array}{l}
R_{l} = \\
\hspace{-0.2cm} \mathop {\sup }\limits_{ \Pr [{\mathbf{u}}_l\left| {\mathbf{s}_l} \right.]}
\left\{  { \left( I\left( {{\mathbf{u}}_l;{\mathbf{y}_l},{\mathbf{H}_l}} \right) - I\left( {{\mathbf{u}_l};{\mathbf{s}_l}} \right) \right)| { f_l({\mathbf{u}_l,\mathbf{s}_l})  = {\mathbf{u}_l} - \mathbf{F}_l \mathbf{s}_l} } \right\}.
\end{array}
\end{equation}
Given any fixed linear assignment matrix $\mathbf{F}_l$, it was proved in \cite[Theorem 1]{Bennatan}  that the maximum rate in (\ref{fading_rate_la}) is achieved by choosing
$\mathbf{x}_l$ to be jointly Gaussian with $\mathbf{s}_l$, which in turn determines
the conditional distribution $\Pr [{\mathbf{u}_l}\left| {\mathbf{s}_l} \right.]$ as follows \cite{Bennatan}
\begin{equation}\label{pr_pdf}
\begin{array}{l}
\Pr \left[ {{\bf{u}}_l \left| {{\bf{s}}_l } \right.} \right] = \\
 \qquad \frac{1}{{\sqrt {\det \left( {\pi {\boldsymbol{\Sigma }}_{{\bf{u}}_l ,{\bf{s}}_l } } \right)} }}\exp \left( { - \left( {{\bf{u}}_l  - {\mathbf{J}_l \bf{s}}_l } \right)^H {\boldsymbol{\Sigma }}_{{\bf{u}}_l,{\bf{s}}_l }^{ - 1} \left( {{\bf{u}}_l  - { \mathbf{J}_l \bf{s}}_l } \right)} \right)
\end{array}
\end{equation}
where
\begin{equation}\label{J}
\mathbf{J}_l = \left( {{ \mathbf{F}_l \bf{\Sigma }}_{{\bf{s}}_l }  + {\boldsymbol{\Sigma }}_{{\bf{x}}_l ,{\bf{s}}_l }^H } \right){\boldsymbol{\Sigma }}_{{\bf{s}}_l }^{ - 1}
\end{equation}
and
\begin{equation}\label{achievable_rate_add}
\begin{array}{l}
{\bf{\Sigma }}_{{\bf{u}}_l, {{\bf{s}}_l }}  = \left( {{\mathbf{F}_l\bf{\Sigma }}_{{\bf{s}}_l } \mathbf{F}_l^H  + {\mathbf{F}_l\bf{\Sigma }}_{{\bf{x}}_l,{\bf{s}}_l }  + {\bf{\Sigma }}_{{\bf{x}}_l,{\bf{s}}_l }^H \mathbf{F}_l^H  + {\bf{\Sigma }}_{{\bf{x}}_l } } \right) \\
 \hspace{2cm} - \left( {\mathbf{F}_l{\bf{\Sigma }}_{{\bf{s}}_l }  + {\bf{\Sigma }}_{{\bf{x}}_l,{\bf{s}}_l }^H } \right){\bf{\Sigma }}_{{\bf{s}}_l }^{ - 1} \left( {\mathbf{F}_l {\bf{{\boldsymbol{\Sigma}} }}_{{\bf{s}}_l }  + {\bf{\Sigma }}_{{\bf{x}}_l,{\bf{s}}_l }^H } \right)^H
\end{array}
\end{equation}
in which
${\bf{\Sigma}}_{{\bf{x}}_l }$, ${\bf{ \Sigma}}_{{\bf{s}}_l}$, and
${\boldsymbol{\Sigma}} _{{\bf{x}}_l,{\bf{s}}_l }$ denotes the covariance matrix of $\mathbf{x}_l$,
the covariance matrix of $\mathbf{s}_l$, and the cross-covariance matrix of $\mathbf{x}_l$ and $\mathbf{s}_l$, respectively.
Then, the  achievable transmission rate in (\ref{fading_rate}) under the linear assignment operation  $f_l\left( {\mathbf{u}_l} , {\mathbf{s}_l} \right) =  \mathbf{u}_l - \mathbf{F}_l \mathbf{s}_l$
and the corresponding conditional distribution in (\ref{pr_pdf}) is defined as the linear assignment capacity \cite{Bennatan}.
Furthermore, it was proved in \cite[Theorem 2]{Bennatan} that the linear assignment capacity region of model (\ref{model}) is found by
choosing the transmitted vector $\mathbf{x}_l$ independent\footnote{It should be noted here we do not require that for
a given user $l$, the maximum value in (\ref{fading_rate})
must choose  $\mathbf{x}_l$ to be independent of $\mathbf{s}_l$. Instead, the conclusion in \cite[Theorem 2]{Bennatan} indicates
that by replacing  the entire given signal set $\mathbf{x}_l$, $l = 1,2,\cdots,L$, which may not be independent of
the entire interference set $\mathbf{s}_l$, $l = 1,2,\cdots,L$ (corresponding to some set of
points on the linear assignment capacity region), to two independent sets
$\mathbf{\widetilde{x}}_l$, $l = 1,2,\cdots,L$ and $\mathbf{\widetilde{s}}_l$, $l = 1,2,\cdots,L$,
the entire linear assignment capacity region can be exhausted. For details, refer to  Appendix V in \cite{Bennatan}. }
of the interference vector $\mathbf{s}_l$, $l= 1,2,\cdots,L$.
In this case, we have ${\bf{\Sigma}}_{{\bf{x}}_l } = {\bf{\Sigma}} _l$, ${\bf{ \Sigma}}_{{\bf{s}}_l}  = \mathbf{\Sigma}_{S,\,l} =
\sum\nolimits_{t = 1}^{l-1}{\bf{\Sigma}}_t $, ${\boldsymbol{\Sigma}} _{{\bf{x}}_l,{\bf{s}}_l} = \mathbf{0}$,
and the covariance matrix of the transmitted signal ${\bf{x}}_T$ can be expressed as
 ${\boldsymbol{\Sigma }}_T   = E\left[ {{\bf{x}}_T{\bf{x}}_T^H } \right] =  \sum\nolimits_{l = 1}^L {{\boldsymbol{\Sigma}}_l }$.
The transmitter is subject to an average power constraint $P$, which implies ${\rm{tr}}\left( {{\boldsymbol{\Sigma }}_T } \right) \le P$.
Also, the conditional distribution
$\Pr [{\mathbf{u}_l}\left| {\mathbf{s}_l} \right.]$ in (\ref{pr_pdf}) reduces to
\begin{equation}\label{pr_pdf_2}
\begin{array}{l}
\Pr \left[ {{\bf{u}}_l \left| {{\bf{s}}_l } \right.} \right] = \\
\quad \frac{1}{{\sqrt {\det \left( {\pi {\boldsymbol{\Sigma}}_l } \right)} }}\exp \left( { - \left( {{\bf{u}}_l  - {\mathbf{F}_l \bf{s}}_l } \right)^H {\boldsymbol{\Sigma}}_l^{ - 1} \left( {{\bf{u}}_l  - {\mathbf{F}_l \bf{s}}_l } \right)} \right).
\end{array}
\end{equation}

Then, by adapting the LAAR expression in \cite[Eq. (35)]{Bennatan} to  complex channels,
the LAAR of the $l$-th receiver in (\ref{fading_rate_la}) can be written as
\begin{equation}\label{achievable_rate_2}
R_{l} = \log_2 \det \left( {{\bf{\Sigma}} _{{\mathbf{u}_l}, {\mathbf{s}_l}} } \right)
- E \left[ {\log_2 \det \left( {\bf{\Sigma}} _{{\mathbf{u}_l},\, {\mathbf{y}_l} \left|{ {\mathbf{H}_l}} \right.} \left( {\mathbf{H}_l} \right)\right)} \right] \\
\end{equation}
where
\begin{equation}\label{achievable_rate_add_2}
\begin{array}{l}
 {\bf{\Sigma }}_{{\bf{u}}_l, \,{\bf{y}}_l \left| {{\bf{H}}_l }\right.} \left( {{\bf{H}}_l } \right) \!=\! \left( { \mathbf{F}_l {\boldsymbol{\Sigma}} _{{\bf{s}}_l } \mathbf{F}_l^H \! + \! \mathbf{F}_l {\boldsymbol{\Sigma}} _{{\bf{x}}_l , {\bf{s}}_l } \! + \! {\boldsymbol{\Sigma}} _{{\bf{x}}_l, {\bf{s}}_l }^H \mathbf{F}_l^H \! + \! {\boldsymbol{\Sigma}} _{{\bf{x}}_l } } \right) \\
\hspace{1cm} - \left[ {\left( {\mathbf{F}_l {\boldsymbol{\Sigma}} _{{\bf{s}}_l }  + \mathbf{F}_l {\boldsymbol{\Sigma}}_{{\bf{x}}_l ,{\bf{s}}_l } \! + \! {\boldsymbol{\Sigma}}_{{\bf{x}}_l ,{\bf{s}}_l }^H \! + \! {\boldsymbol{\Sigma}} _{{\bf{x}}_l } } \right){\bf{H}}_l^H } \right] \\
 \times \left[ {{\bf{H}}_l \left( {{\boldsymbol{\Sigma}}_{{\bf{s}}_l }  + {\boldsymbol{\Sigma}} _{{\bf{x}}_l }  + {\boldsymbol{\Sigma}} _{{\bf{x}}_l ,{\bf{s}}_l }  + {\boldsymbol{\Sigma}} _{{\bf{x}}_l, {\bf{s}}_l }^H } \right){\bf{H}}_l^H \! + \! {\boldsymbol{\Sigma}} _{Z,l} \left( {{\bf{H}}_l } \right)} \right]^{ - 1} \! \\
\hspace{0.5cm}  \times \left[ {\left( \mathbf{F}_l {\boldsymbol{\Sigma}} _{{\bf{s}}_l }  + \mathbf{F}_l{\boldsymbol{\Sigma}} _{{\bf{x}}_l, {\bf{s}}_l }  + {\boldsymbol{\Sigma}} _{{\bf{x}}_l , {\bf{s}}_l }^H  + {\boldsymbol{\Sigma}} _{{\bf{x}}_l } \right){\bf{H}}_l^H } \right]^H  \\
 \end{array}
\end{equation}
in which $\mathbf{\Sigma}_{Z,\, l}(\mathbf{H}_l) =\mathbf{H}_l \left(  \sum\nolimits_{t = l+1}^{L}{\bf{\Sigma}}_t \right)\mathbf{H}_l^H + N_0 \mathbf{I}_{N_r}$.
Since $\mathbf{x}_l$ and $\mathbf{s}_l$ are independent (${\boldsymbol{\Sigma}} _{{\bf{x}}_l, {\bf{s}}_l} = \mathbf{0}$),
(\ref{achievable_rate_add}) and (\ref{achievable_rate_add_2}) can be simplified as follows
\begin{equation}\label{achievable_rate_3}
{\boldsymbol{\Sigma}} _{{\mathbf{u}_l}, {\mathbf{s}_l}}  = {\bf{\Sigma}} _l
\end{equation}
\begin{equation}\label{achievable_rate_4}
\begin{array}{l}
 {\bf{\Sigma}} _{{\mathbf{u}_l},\,{\mathbf{y}_l}\left| {{\mathbf{H}_l}} \right.} \left( {\mathbf{H}_l} \right) =  \\
\hspace{1cm} {\mathbf{C}_l} - {\mathbf{A}_l \mathbf{H}_l^H}
  \left[ {{\mathbf{H}_l \mathbf{B}_l {\mathbf{H}_l^H}}  +  \mathbf{\Sigma}_{Z,\, l} (\mathbf{H}_l)  } \right]^{ - 1} {\mathbf{H}_l} \mathbf{A}_l^H  \\
 \end{array}
\end{equation}
where ${\bf{A}}_l = {\bf{F}}_l\mathbf{\Sigma}_{S,\,l}  + {\bf{\Sigma}}_l$,
${\bf{B}}_l = \mathbf{\Sigma}_{S,\,l} + {\bf{\Sigma}}_l$, and ${\bf{C}}_l = {{\bf{F}}_l \mathbf{\Sigma}_{S,\,l} {\bf{F}}_l^H }  + {\bf{\Sigma}}_l $.

{
Here we present the encoding and decoding process to achieve the linear assignment achievable rate in (11).

Encoding process at the transmitter:

\begin{enumerate}

\item[1)]  First select a transmitted signal $\mathbf{x}_1$ for receiver 1.

 \item[2)]  Generate $e^{nI(\mathbf{u}_2;\mathbf{y}_2,\mathbf{H}_2)}$ independent sequences.

 \item[3)]  Distribute these sequences into the $e^{nR_2}$ bin codebook uniformly.

 \item[4)]  Given the non-causally known interference $\mathbf{s}_2 = \mathbf{x}_1$ and the message $W_2 = k$ for receiver 2, look
 for a joint typical pair \cite{Cover} ($\mathbf{u}_2$, $\mathbf{s}_2$)  among the sequences in bin $k$.

 \item[5)]  The signal ${\bf{x}}_{2}$ for receiver  $2$ is constructed using the linear assignment operation $\mathbf{x}_2 = \mathbf{u}_2 - \mathbf{F}_2 \mathbf{s}_2$.

 \item[6)]  The signal $\mathbf{x}_l$ for receiver $l$ is constructed in a similar manner above. The signal ${\bf{s}}_{ l }  = {\bf{x}}_{  1 }  + {\bf{x}}_{  2 }  +  \cdots + {\bf{x}}_{  {l - 1}}$ is regarded as non-causally known interference. This process continues to the $L$-th receiver.

\end{enumerate}

Decoding process at receiver $l$:

\begin{enumerate}
\item[1)] For the received signal $\mathbf{y}_l$,  look for a joint typical pair  ($\mathbf{u}_l$, $\mathbf{y}_l$, $\mathbf{H}_l$)
among the sequences in the codebook.

\item[2)] Declare an error when more than one joint typical pairs ($\mathbf{u}_l$, $\mathbf{y}_l$, $\mathbf{H}_l$)
are found. Also, declare an error when no joint typical pair ($\mathbf{u}_l$, $\mathbf{y}_l$, $\mathbf{H}_l$) is found.

\item[3)] Set the estimate $\widehat{W}_l$ equal to the index of the bin containing this sequence $\mathbf{u}_l$.

\end{enumerate}

}

With the LAAR in (\ref{achievable_rate_2}),  the LAWSR of  model (\ref{model}) is given by
\begin{equation}\label{wsum}
R_{\rm{sum}}^{\rm{w}}  = \sum\limits_{l = 1}^L {\mu _l R_{l} }
\end{equation}
where $R_{l}$ is evaluated in (\ref{achievable_rate_2}) and $\mu _l$, $ l = 1,2,\cdots,L$ are nonnegative
weights\footnote{In a practical communication system, these weights can be determined according to different service requirements for different users.} satisfying
$\sum\nolimits_{l = 1}^L {\mu _l } = L$.  The sequel resorts to develop a transmit design of $\mathbf{F}_l$ and $\mathbf{\Sigma}_l$, $l = 1,2,\cdots,L$,
under the constraint ${\rm{tr}} \left( \sum\nolimits_{l = 1}^L {{\bf{\Sigma}}_l } \right) \le P$, by maximizing the LAWSR in (\ref{wsum}).

\section{Transmit Design for Maximizing the LAWSR}

In this section, we investigate the transmit design to maximize the LAWSR in (\ref{wsum}).
We begin by establishing  necessary conditions that the optimal $\mathbf{F}_l$ and $\mathbf{\Sigma}_l$ must satisfy.
Then, an  algorithm is developed to optimize $\mathbf{F}_l$ and $\mathbf{\Sigma}_l$ iteratively.

\subsection{Necessary Conditions for the Optimal Design}
From (\ref{wsum}),  we know that the objective function $R_{\rm{sum}}^{\rm{w}}$ is a non-convex function of the matrices
$\mathbf{F}_l$ and $\mathbf{\Sigma}_l$,  $l = 1,2,\cdots,L$.  Thus, we obtain a set of necessary conditions for the
optimal linear assignment matrices and precoding matrices below.
\begin{theorem}\label{nec_cond}
The optimal transmit design, which maximizes the LAWSR in (\ref{wsum}), satisfies the following
conditions:
{ \begin{equation}\label{cond_1}
\begin{array}{l}
\hspace{-0.2cm} - \mu_l \log _2 e E_{{\bf{H}}_l } \left[ {\boldsymbol{\Sigma} _{{\bf{u}}_l \left| {{\bf{y}}_l , \, {\bf{H}}_l } \right.} \left( {{\bf{H}}_l } \right)^{ - 1} \left({\bf{F}}_l   - {\bf{T}}_l \left( {{\bf{H}}_l } \right) \right)} {\boldsymbol{\Sigma }}_{S,\, l} \right] = \mathbf{0}, \\
  \hspace{6cm} l = 1,2,\cdots,L
 \end{array}
\end{equation} }
\begin{equation}\label{cond_2}
\boldsymbol \Sigma _l  = {\bf{P}}_l {\bf{P}}_l^H,  \quad l = 1,2,\cdots,L
\end{equation}
{ \begin{equation}\label{cond_3}
\begin{array}{l}
 \hspace{-0.2cm} \left( {{\bf{P}}_t^H } \right)^{ - 1} \! - \! \log _2 e \, \left[\sum\limits_{l = 1}^{t - 1} \mu_l {E_{{\bf{H}}_l } \left[{\bf{G}}_{1, \, l} \left( {{\bf{H}}_l } \right)\right] + \mu_t E_{{\bf{H}}_t } \left[ {{\bf{G}}_{2, \, t} \left( {{\bf{H}}_t } \right)} \right]} \right.\\
\hspace{3cm}  \left.{ + \sum\limits_{l = t + 1}^L \mu_l {E_{{\bf{H}}_l } [{\bf{G}}_{3, \, l} \left( {{\bf{H}}_l } \right)]} } \right]{\bf{P}}_t = \theta{\bf{P}}_t,  \\
\hspace{6cm}  t = 1,2,  \cdots  ,L
 \end{array}
\end{equation} }
\begin{equation}\label{cond_4}
\theta \left( {\sum\limits_{l = 1}^L {{\rm{tr}}\left( {{\mathbf{P}}_l {\mathbf{P}}_l^H } \right) - P} } \right) = 0
\end{equation}
\begin{equation}\label{cond_5}
{\sum\limits_{l = 1}^L {{\rm{tr}}\left( {{\mathbf{P}}_l {\mathbf{P}}_l^H } \right) - P} } \leq 0
\end{equation}
\begin{equation}\label{cond_6}
\theta \geq 0.
\end{equation}
 ${\bf{G}}_{1,\, l} \left( {{\bf{H}}_l } \right)$, ${\bf{G}}_{2,\, t} \left( {{\bf{H}}_t } \right)$, and
${\bf{G}}_{3,\,l} \left( {{\bf{H}}_l } \right)$ are $N_t \times N_t$ matrices given by
\begin{equation}\label{G1_H}
{\bf{G}}_{1, \, l} \left( {{\bf{H}}_l } \right) = {\bf{T}}_l^H \left( {{\bf{H}}_l } \right)\boldsymbol \Sigma _{{\bf{u}}_l \left| {{\bf{y}}_l , \, {\bf{H}}_l } \right.} \left( {{\bf{H}}_l } \right)^{ - 1} {\bf{T}}_l \left( {{\bf{H}}_l } \right)
\end{equation}
\begin{equation}\label{G2_H}
\begin{array}{l}
 {\bf{G}}_{2,\, t} \left( {{\bf{H}}_t } \right) \! = \! \boldsymbol \Sigma _{{\bf{u}}_t \left| {{\bf{y}}_t ,\, {\bf{H}}_t } \right.} \left( {{\bf{H}}_t } \right)^{ - 1} \! - \! {\bf{T}}_t \left( {{\bf{H}}_t } \right) \boldsymbol\Sigma _{{\bf{u}}_t \left| {{\bf{y}}_t , \, {\bf{H}}_t } \right.} \left( {{\bf{H}}_l } \right)^{ - 1}   \\
   \hspace{2cm} + {\bf{T}}_t^H \left( {{\bf{H}}_t } \right) \boldsymbol\Sigma _{{\bf{u}}_t \left| {{\bf{y}}_t , \, {\bf{H}}_t } \right.} \left( {{\bf{H}}_t } \right)^{ - 1} {\bf{T}}_t \left( {{\bf{H}}_t } \right) \\
 \hspace{4cm}  - \boldsymbol\Sigma _{{\bf{u}}_t \left| {{\bf{y}}_t , \, {\bf{H}}_t } \right.} \left( {{\bf{H}}_t } \right)^{ - 1} {\bf{T}}_t \left( {{\bf{H}}_t } \right) \\
 \end{array}
\end{equation}
\begin{equation}\label{G3_H}
\begin{array}{l}
 {\bf{G}}_{3,\, l} \left( {{\bf{H}}_l } \right) = {\bf{F}}_l^H \boldsymbol \Sigma _{{\bf{u}}_l \left| {{\bf{y}}_l , \, {\bf{H}}_l } \right.} \left( {{\bf{H}}_l } \right)^{ - 1} {\bf{F}}_l  - {\bf{T}}_l^H \left( {{\bf{H}}_l } \right)   \\
 \times\boldsymbol\Sigma _{{\bf{u}}_l \left| {{\bf{y}}_l , \, {\bf{H}}_l } \right.} \left( {{\bf{H}}_l } \right)^{ - 1} {\bf{F}}_l \! + \! {\bf{T}}_l^H \left( {{\bf{H}}_l } \right)\boldsymbol\Sigma _{{\bf{u}}_l \left| {{\bf{y}}_l , \, {\bf{H}}_l } \right.} \! \left( {{\bf{H}}_l } \right)^{ - 1} \! {\bf{T}}_l \left( {{\bf{H}}_l } \right) \\
 \hspace{3cm} - {\bf{F}}_l^H \boldsymbol\Sigma _{{\bf{u}}_l \left| {{\bf{y}}_l , \, {\bf{H}}_l } \right.} \left( {{\bf{H}}_l } \right)^{ - 1} {\bf{T}}_l \left( {{\bf{H}}_l } \right) \\
 \end{array}
\end{equation}
where
\begin{equation}\label{T1_H}
{\bf{T}}_l \left( {{\bf{H}}_l } \right) = {\bf{A}}_l {\bf{H}}_l^H \left[ {{\bf{H}}_l {\bf{B}}_l {\bf{H}}_l^H + {\bf{\Sigma }}_{Z,\, l} } \right]^{ - 1} {\bf{H}}_l .
\end{equation}
\begin{proof}
See Appendix \ref{proof:nec_cond}.
\end{proof}
\end{theorem}

Equations (\ref{cond_1})--(\ref{cond_6}) provide elementary conditions that characterize the optimal  designs of $\mathbf{F}_l, \mathbf{P}_l, l = 1,2,\cdots,L$.
In general, finding  closed-form expressions for the optimal  designs from Theorem \ref{nec_cond} is a difficult task,
if not intractable. The problems are complex because of their non-convexity
and highly involved representation.  Indeed, the expectation operation $E \left[\cdot \right]$ in (\ref{cond_1}) and
(\ref{cond_3}) requires averaging of all possible realizations of channel matrices $\mathbf{H}_l$.   The inverse operation $(\cdot)^{-1}$
results in an involved structure of  $\mathbf{F}_l, \mathbf{P}_l$ in (\ref{cond_1}) and (\ref{cond_3}).
 These pose serious challenges. { Nevertheless, Theorem \ref{nec_cond} provides gradient descent directions of the LAWSR in (\ref{wsum})
 with respect to $\mathbf{F}_l$ and $\mathbf{P}_l$, from which a numerical algorithm can be formulated
 to search for the optimal  designs iteratively. }

\subsection{Iterative Algorithm for LAWSR Maximization}

From (\ref{cond_1}) and (\ref{cond_3}), it can be seen that the optimal matrices $\mathbf{F}_l$ and $\mathbf{P}_l$ depend on one another, which
leads to a multidimensional optimization problem.  We resort to a prevalent approach in dealing with this type of problem, in terms of iteratively optimizing
 one variable at a time with others fixed. { Within each iteration, we exploit the gradient descent update
 via the partial  derivatives of the LAWSR in (\ref{wsum}) with respect to $\mathbf{F}_l$ and $\mathbf{P}_l$.  These partial derivatives $\nabla _{{\bf{F}}_l } R_{{\rm{sum}}}^{\rm{w}}$
 and $\nabla _{{\bf{P}}_l } R_{{\rm{sum}}}^{\rm{w}}$ are specified by the left-hand terms of (\ref{cond_1}) and  (\ref{cond_3}) in Theorem 1, respectively. } Combining this search direction with the backtracking line
 search conditions \cite{Boyd}, Algorithm 1 maximizes the LAWSR over $\mathbf{F}_l$ and $\mathbf{P}_l$.

{
\begin{alg} \label{Gradient_FBC}

{Maximizing the LAWSR over $\mathbf{F}_l$ and $\mathbf{P}_l$.}

\vspace*{2.5mm} \hrule \vspace*{1mm}

  \begin{enumerate}
\item   Initialize $\mathbf{F}_l^{(1)}$ and $\mathbf{P}_l^{(1)}$, $l = 1,2,\cdots,L$. Set $n = 1$,
  the tolerance for the backtracking line search  $\varepsilon_1 >0$,
the tolerance for stoping the algorithm $\varepsilon_2 >0$,  and the maximum iteration number
  $N_{\rm max} $.  Select values for the backtracking line search  parameter $\beta$ with $\beta  \in (0,1)$.

\item   Compute $R_{\rm{sum}}^{\rm{w},(n)}$ and $\nabla _{{\bf{F}}_l } R_{{\rm{sum}}}^{\rm{w}}$,
based on $\mathbf{F}_l^{(n)}$ and  $\mathbf{P}_l^{(n)}$,
  $l = 1,2,\cdots,L$.

\item Initialize step size $t = 1$.

\item If $t < \varepsilon_1 $, then go to step 8.

\item Compute $\mathbf{{F}}_l^{(n + 1)} = \mathbf{F}_l^{(n)} + t  \nabla _{{\bf{F}}_l } R_{{\rm{sum}}}^{\rm{w}} $, $l = 1,2,\cdots,L$.

\item Evaluate ${R}_{\rm{sum}}^{\rm{w},(n + 1)}$,  based on $\mathbf{F}_l^{(n + 1)}$ and  $\mathbf{P}_l^{(n)}$, $l = 1,2,\cdots,L$.

\item Set $t := \beta t $.  If ${R}_{\rm{sum}}^{\rm{w},(n + 1)} < R_{\rm{sum}}^{\rm{w},(n)} $, go to step 4.

\item  Compute $\nabla _{{\bf{P}}_l } R_{{\rm{sum}}}^{\rm{w}}$, based on $\mathbf{F}_l^{(n + 1)}$ and  $\mathbf{P}_l^{(n)}$,
 $l = 1,2,\cdots,L$.

\item Initialize step size $u = 1$.

\item If $u < \varepsilon_1$, then go to step 15.

\item   Compute $\mathbf{{P}}_l^{(n + 1)} = \mathbf{P}_l^{(n)} + u  \nabla _{{\bf{P}}_l } R_{{\rm{sum}}}^{\rm{w}} $, $l = 1,2,\cdots,L$.

\item   If $\sum\nolimits_{l = 1}^L {\tr\left( {{\mathbf{P}}_l^{\left( n + 1 \right)}}\left( {\mathbf{P}}_l^{\left( n + 1\right)}\right)^H  \right)} > P$, update ${\mathbf{P}}_l^{\left( n + 1\right)}  :  = \frac{{\sqrt {P } {\mathbf{P}}_l^{\left( n + 1\right)} }}{{\left[ {\sum\nolimits_{l = 1}^L {\left\| {{\mathbf{P}}_l^{\left( n + 1\right)} } \right\|_F^2 } } \right]^{1/2} }}$, $l = 1,2,\cdots, L$.

\item  Evaluate ${R}_{\rm{sum}}^{\rm{w},(n + 1)}$,  based on $\mathbf{{F}}_l^{(n + 1)}$ and  $\mathbf{{P}}_l^{(n + 1)}$,
  $l = 1,2,\cdots,L$.

\item Set $u := \beta u $.  If ${R}_{\rm{sum}}^{\rm{w},(n + 1)} < R_{\rm{sum}}^{\rm{w},(n )} $, go to step 10.

\item  If $R_{\rm{sum}}^{\rm{w},(n + 1)} - {R}_{\rm{sum}}^{\rm{w},(n)} > \varepsilon_2$ and $n < N_{\rm max}$, set $n : = n + 1$, go to step 2;
otherwise, stop the algorithm.

\vspace*{1.5mm}
\hrule

\vspace*{1.5mm}

  \end{enumerate}

\end{alg}
\null
\par

}
{
Since it is generally very difficult to obtain a closed-form expression for $R_l$ in (\ref{achievable_rate_2}), we employ Monte Carlo simulation to estimate the value of $R_{\rm{sum}}^{\rm{w}} $ and
the gradients $\nabla _{{\bf{F}}_l } R_{{\rm{sum}}}^{\rm{w}}$ and $\nabla _{{\bf{P}}_l } R_{{\rm{sum}}}^{\rm{w}}$ in Algorithm 1.
Such an approach has been also used in \cite{Wang2011,Vaze2009,Wu2012TVT,Wu2013TCOM}. }

\section{Low Complexity Design Based on Second-Order Statistics of the CSI}
It should be noted that the main drawback of Algorithm 1 is the exhaustive averaging
in each iteration, which might cause long execution time. Therefore,
in this section, we propose a low complexity transmit design over the fading MIMO BC in (\ref{model})
based on second-order statistics of the CSI.
Before addressing this, we first derive an upper bound on the LAAR for each receiver in (\ref{achievable_rate_2}). The derived upper bound admits a highly efficient searching algorithm for
the linear assignment matrices and the precoding matrices.  In addition, we investigate the precoding strategies in two special cases.

\subsection{An Upper Bound of the Linear Assignment Achievable Rate}
In the following theorem, we provide an upper bound of the linear assignment achievable rate for each receiver in (\ref{achievable_rate_2}).
\begin{theorem}\label{upper_bound}
The linear assignment achievable rate of the $l$-th receiver of the fading MIMO BC model, given in (\ref{achievable_rate_2}), can be upper bounded by
\begin{equation}\label{bound_prop}
\begin{array}{l}
 R_l  \le R_{{\rm{upp}},\,l} = \log _2 \det \left( {{\bf{\Sigma }}_l } \right) - \log _2 \det \left( {{\bf{C}}_l } \right) \\
  -  \left[\log _2 \det \left( {{\bf{R}}_{g,\,l} \left( {{\bf{D}}_l  + \sum\limits_{t = l + 1}^L {{\bf{\Sigma }}_t } } \right) + N_0 {\bf{I}}_{N_t } } \right) \right. \\
 \left.  - \log _2 \det \left( {{\bf{R}}_{g,\,l} \left( {{\bf{B}}_l  + \sum\limits_{t = l + 1}^L {{\bf{\Sigma }}_t } } \right) + N_0 {\bf{I}}_{N_t } } \right) \right] \\
 \end{array}
\end{equation}
where $\mathbf{D}_l = \mathbf{B}_l - \mathbf{A}_l^H \mathbf{C}_l^{-1} \mathbf{A}_l$, and
$\mathbf{R}_{g,\, l} = E \left[\mathbf{H}_l^H \mathbf{H}_l \right] \succ \mathbf{0}$ reflects
the second-order statistics of the fading channel $\mathbf{H}_l$.
\begin{proof}
See Appendix \ref{proof:upper_bound}.
\end{proof}
\end{theorem}

Here, Jensen's inequality, a common tool in  MIMO capacity analysis and for deriving power allocation strategies \cite{Li,Li_2,Sharif,Somekh,Goel}, is applied to obtain the upper bound
in (\ref{bound_prop}).  Specifically,  we apply  Jensen's inequality to
two terms in (\ref{Prop_proof_5}) in Appendix \ref{proof:upper_bound} and then subtract them.  Since the ``bounding errors" for both terms are similar, subtracting these will have a canceling effect  which in turn will yield a fairly tight upper bound. This is confirmed in numerical results where the transmit design in the context of
the derived upper bound performs close to the design via Algorithm 1  based on the exact LAAR in (\ref{achievable_rate_2}).

{
It should be noted that employing Jensen's inequality to two subtraction terms
simultaneously is a useful technique for the transmission design in fading channels, whilst
having applicability to  problems such as those relating to the MIMO BC, the MIMO interference channels,
and massive MIMO systems \cite{Li_3,Li_4,Adhikary2013TIT,Liang2014}. To the best of our knowledge, Theorem
2 reveals for the first time that this type of subtraction results in a strict theoretical upper bound of the
original problem, instead of simply an approximation. }

\subsection{A Low Complexity Design}
We provide a closed-form design of the linear assignment matrix $\mathbf{F}_l$ based on the upper bound in (\ref{bound_prop}) as follows.
\begin{proposition}\label{upper_bound_F}
For a random fading channel $\mathbf{H}_l$ satisfying  $E \left[\mathbf{H}_l^H \mathbf{H}_l \right] = \mathbf{R}_{g,\, l}$, a  closed-form solution
for the linear assignment matrix  $\widetilde{{\bf{F}}}_l$,  which maximizes the upper bound in (\ref{bound_prop}), is given by
\begin{equation}\label{PI_F}
{\widetilde{{\bf{F}}}}_l  = {\bf{\Sigma}} _l \left( {{\bf{\Sigma}}_l  + \sum\limits_{t = l + 1 }^{L} {{\bf{\Sigma }}_t}  +  N_0 {\bf{R}}_{g,\,l }^{ - 1} } \right)^{ - 1}, \quad l = 1,2,\cdots,L.
\end{equation}
The achievable rate of the $l$-th receiver $\widetilde{R}_l$, under the linear assignment matrix design in (\ref{PI_F}), can
be upper bounded\footnote{It is noted that here we evaluate the exact achievable rate $R_l$ by substituting ${\widetilde{{\bf{F}}}}_l$ in (\ref{PI_F}) into (\ref{achievable_rate_2}).} by
\begin{equation}\label{R_upper_F}
\begin{array}{l}
\widetilde{R}_l  \le \widetilde{R}_{{\rm{upp}},\,l}
 = \log _2 \det \left( {{\bf{R}}_{g,\,l} \sum\limits_{t = l}^L {{\bf{\Sigma }}_t }  + N_0 {\bf{I}}_{N_t } } \right) \\
 \hspace{3cm} - \log _2 \det \left( {{\bf{R}}_{g,\,l} \sum\limits_{t = l + 1}^L {{\bf{\Sigma }}_t }  + N_0 {\bf{I}}_{N_t } } \right).
  \end{array}
\end{equation}
\begin{proof}
See Appendix \ref{proof:upper_bound_F}.
\end{proof}
\end{proposition}

Interestingly, equation (\ref{PI_F}) demonstrates that the linear assignment matrix ${\widetilde{{\bf{F}}}}_l$ can be designed simply by exploiting
the second-order statistics of the CSI, whereas the matrix structure is similar to that of DPC, designed via instantaneous CSI in \cite{Bennatan}.
Moreover, the structure of  $\widetilde{R}_{{\rm{upp}},\,l}$ in (\ref{R_upper_F}) is similar to  dirty-paper transmission rate
\cite[Eq. (2.18)]{Jindal} with instantaneous CSI, where  the impact of non-causally known interference
 $\sum\nolimits_{t = 1}^{l - 1}{{\bf{\Sigma}} _{t}} $ in (\ref{achievable_rate_2}) does not exist.  Henceforth,
 various highly efficient algorithms  for the MIMO BC model with instantaneous CSI can be utilized
to design the  matrices ${\bf{\Sigma }}_l$. To summarize, we present an algorithm with  reduced computational complexity to design $\mathbf{F}_l$ and
$\mathbf{\mathbf{\Sigma}}_l$ as follows.

\begin{alg} \label{Gradient_FBC_low}

\vspace*{3mm}
\hrule

\vspace*{1.5mm}
{A low complexity transmit design over the fading MIMO BC.}
  \begin{enumerate}
\vspace*{1.5mm}
\hrule
\vspace*{1.5mm}

 \item Find ${\bf{{\Sigma} }}_l$, $l = 1,2,\cdots,L$, maximizing  $\widetilde R_{\rm upp,\,sum}^w  = \sum\nolimits_{l = 1}^L {\mu_l \widetilde R_{{\rm{upp}},\, l} }$ by classical algorithms of the conventional  MIMO BC model with instantaneous CSI such as in \cite{Jindal,Christensen2008TWC}.

\item Design ${\widetilde{{\bf{F}}}}_l$, $l = 1,2,\cdots,L$,  as in (\ref{PI_F}) via the obtained  ${\bf{{\Sigma} }}_l$.

  \end{enumerate}

\vspace*{1.5mm}
\hrule

\vspace*{1.5mm}

\end{alg}
\null
\par

\subsection{Transmit Strategies in Two Special Cases}
In the following,  we discuss the transmit strategies in two special cases based on Proposition \ref{upper_bound_F}.

\begin{proposition}\label{upper_bound_one}
For the special case when $N_t = 1$, a near-optimal power allocation strategy is given by
\begin{equation}\label{optimal_power}
\begin{array}{l}
 P_l  = \left\{ \begin{array}{lll}
 P, &{l = \widetilde l }\\
 0, &{\rm{otherwise}}& \\
 \end{array} \right. \\
 \widetilde l = \arg \max_{l} \left\{  r_{ l} \right\}, \quad l = 1,2,\cdots,L \\
 \end{array}
\end{equation}
where  $r_{l} = E \left[\mathbf{h}_l^H \mathbf{h}_l \right]$.
\begin{proof}
See Appendix \ref{proof:upper_bound_one}.
\end{proof}
\end{proposition}

Proposition \ref{upper_bound_one} implies that the multiuser diversity gain is achieved by an opportunistic scheduling scheme. This is similar to results  in \cite{Viswanath}, which applied for the case $N_t = 1$ and with instantaneous CSI at the transmitter.

\begin{proposition}\label{upper_bound_iid}
If all the receivers experience \emph{i.i.d. fading conditions} with zero-mean and unit variance, a near-optimal transmit strategy is to perform time-sharing, where all the receivers are served one at a time in a round-robin fashion.
\begin{proof}
According to (\ref{R_upper_F}), the LAWSR $R_{\rm{sum}}^{\rm w}$ in i.i.d. fading channels is upper bounded by
\begin{equation}\label{sum}
R_{\rm{sum}}^{\rm w} \leq \log_2 \det \left( {N_r {\bf{\Sigma}}_T }  + N_0 {\bf{I}}_{N_t }  \right)
\end{equation}
where we have the power constraint ${\rm{tr}}\left( {{\bf{\Sigma }}_T } \right) \le P$.  It is known from \cite{Telatar} that the maximal value of the right term of (\ref{sum}) is achieved by ${{\bf{\Sigma }}_T } =  P/N_t {\bf{I}}_{N_t}$. Then, $R_{\rm{sum}}^{\rm w}$ is bounded by the single-user rate
\begin{equation}\label{LASR}
R_{\rm{sum}}^{\rm w} \leq N_t \ln \left( { \frac{N_r}{N_t} P + N_0 }  \right) .
\end{equation}
\end{proof}
\end{proposition}

Note that Proposition  \ref{upper_bound_iid} corresponds to the conclusion in \cite[Appendix I]{Bennatan}, because in  i.i.d. fading channels,  no statistical CSI can be exploited by the transmitter.

\section{Numerical Results}
This section illustrates the benefits of the transmit designs and examines the efficacy of the proposed
algorithms by several examples. In all these examples, we consider $L = 2$ and normalize the average energy of the receivers' channels
as $E\left[ {{\rm{tr}}\left( {{\bf{H}}_l {\bf{H}}_l^H } \right)} \right] = N_r N_t$, $l = 1,2$.  The signal-to-noise ratio (SNR) is given by $
{\rm{SNR}} = \frac{{E\left[ {{\rm{tr}}\left( {{\bf{H}}_l {\bf{H}}_l^H } \right)} \right]P}}{{N_r N_t N_0 }} = \frac{P}{N_0}
$. Throughout this section, we set $\mu_1 = \mu_2 = 1$. { For Algorithm 1, we set $\varepsilon_1 = 10^{-3}$, $\varepsilon_2 = 10^{-4}$, and $N_{\rm max} = 60$.
{ Due to the non-convexity of the maximization problem in (\ref{wsum}),  Algorithm 1 may only find
    a local optimum LAWSR.  Meanwhile, the explicit expressions in Proposition 1 may provide us some intuitive instructions on the possible structure
    of the optimal transmission design. In order to  exploit this point and avoid local convergence,
    we combine the initialization based on the design of Algorithm 2 and the multiple random initializations\footnote{ The multiple random initializations were adopted in \cite{Serbetli2004TSP,Dumont,Wang2011}.} together as the final initializations for Algorithm 1.
    Then, we choose the obtained designs that achieve the maximal LAWSR as the final solution.
}

\subsection{Comparison of Computational Complexity}
The computational complexity of Algorithm 1 is mainly due to the Monte-Carlo
estimation of the expectation in $R_{\rm{sum}}^{\rm{w}}$,
$\nabla _{{\bf{F}}_l } R_{{\rm{sum}}}^{\rm{w}}$, and $\nabla _{{\bf{P}}_l } R_{{\rm{sum}}}^{\rm{w}}$.
For Algorithm 2, $\mathbf{F}_l$ is designed by a closed-form expression in (\ref{PI_F}). Therefore, the main
computational complexity of Algorithm 2 is the calculation of
$\widetilde R_{\rm upp,\,sum}^w$ and $\nabla _{{\bf{P}}_l } \widetilde R_{\rm upp,\,sum}^w$\footnote{ The expression of this gradient can be easily obtained based on (\ref{R_upper_F}) and following the similar steps as
those in Appendix A.}. Here we compare the computational complexity between
 evaluating $R_{\rm{sum}}^{\rm{w}}$, $\nabla _{{\bf{F}}_l } R_{{\rm{sum}}}^{\rm{w}}$, and $\nabla _{{\bf{P}}_l } R_{{\rm{sum}}}^{\rm{w}}$
in Algorithm 1 and evaluating
 $\widetilde R_{\rm upp,\,sum}^w$ and $\nabla _{{\bf{P}}_l } \widetilde R_{\rm upp,\,sum}^w$ in Algorithm 2.

The accuracy of Monte Carlo estimation can be improved by increasing the number of channel realizations. However, this will also increase
the computational complexity.  Here we provide a method to find a reasonable number of channel realizations. We define the following sets: 1) the number of channel realizations
set $[k,2k,3k,\cdots]$, 2) the estimated $R_{\rm{sum}}^{\rm{w}} $ set $[\widetilde{R}_{\rm{sum}}^{\rm{w},(1)},\widetilde{R}_{\rm{sum}}^{\rm{w},(2)}, \widetilde{R}_{\rm{sum}}^{\rm{w},(3)},\cdots ]$, 3) the estimated $\nabla _{{\bf{F}}_l } {R}_{{\rm{sum}}}^{\rm{w}}$
set $[\nabla _{{\bf{F}}_l } \widetilde{R}_{{\rm{sum}}}^{\rm{w},(1)},\nabla _{{\bf{F}}_l } \widetilde{R}_{{\rm{sum}}}^{\rm{w},(2)},\nabla _{{\bf{F}}_l } \widetilde{R}_{{\rm{sum}}}^{\rm{w},(3)},\cdots ]$,
4) the estimated $\nabla _{{\bf{P}}_l } {R}_{{\rm{sum}}}^{\rm{w}}$ set $[\nabla _{{\bf{P}}_l } \widetilde{R}_{{\rm{sum}}}^{\rm{w},(1)},\nabla _{{\bf{P}}_l } \widetilde{R}_{{\rm{sum}}}^{\rm{w},(2)},
\nabla _{{\bf{P}}_l } \widetilde{R}_{{\rm{sum}}}^{\rm{w},(3)},\cdots]$. When $i k $ channel realizations are used,
$\widetilde{R}_{\rm{sum}}^{\rm{w}, (i)} $, $\nabla _{{\bf{F}}_l } \widetilde{R}_{{\rm{sum}}}^{\rm{w},(i)}$,
and $\nabla _{{\bf{P}}_l } \widetilde{R}_{{\rm{sum}}}^{\rm{w},(i)}$ are the corresponding estimated $R_{\rm{sum}}^{\rm{w}} $,  $\nabla _{{\bf{F}}_l } R_{{\rm{sum}}}^{\rm{w}}$,
and $\nabla _{{\bf{P}}_l } R_{{\rm{sum}}}^{\rm{w}}$, respectively.

We define
\begin{equation}
{i^*} = \arg \mathop {\min }\limits_i \left| {  \widetilde{R}_{\rm{sum}}^{\rm{w},(i + 1)}  - \widetilde{R}_{\rm{sum}}^{\rm{w},( i)} } \right| < \alpha
\end{equation}
for a given threshold $\alpha$. Then, $i^* k$ channel realizations are used to evaluate $R_{\rm{sum}}^{\rm{w}}$.

Similarly, we define
\begin{equation}
{t^*} = \arg \mathop {\min }\limits_t \left| \parallel\nabla _{{\bf{F}}_l } \widetilde{R}_{{\rm{sum}}}^{\rm{w},(t + 1)}\parallel_F  - \parallel\nabla _{{\bf{F}}_l } \widetilde{R}_{{\rm{sum}}}^{\rm{w},(t)}\parallel_F  \right| < \gamma
\end{equation}
\begin{equation}
{p^*} = \arg \mathop {\min }\limits_p \left| \parallel\nabla _{{\bf{P}}_l } \widetilde{R}_{{\rm{sum}}}^{\rm{w},(p + 1)}\parallel_F  - \parallel\nabla _{{\bf{P}}_l } \widetilde{R}_{{\rm{sum}}}^{\rm{w},(p)}\parallel_F  \right| < \gamma
\end{equation}
for a given threshold $\gamma$. Then, $t^* k$ and $p^* k$ channel realizations are used to evaluate $\nabla _{{\bf{F}}_l } R_{{\rm{sum}}}^{\rm{w}}$
and $\nabla _{{\bf{P}}_l } R_{{\rm{sum}}}^{\rm{w}}$, respectively.

Next, we compare the computational complexity of evaluating $R_{\rm{sum}}^{\rm{w}}$,
$\nabla _{{\bf{F}}_l } R_{{\rm{sum}}}^{\rm{w}}$, and $\nabla _{{\bf{P}}_l } R_{{\rm{sum}}}^{\rm{w}}$
in Algorithm 1 and that of evaluating  $\widetilde R_{\rm upp,\,sum}^w$ and $\nabla _{{\bf{P}}_l } \widetilde R_{\rm upp,\,sum}^w$ in Algorithm 2.
We set $k = 1000$, $a = 0.01$, and $\gamma = 0.05$ in our simulation.
The simulations are performed with Matlab on an Intel Core i7-3770 3.4GHz processor. The
running time for evaluating $R_{\rm{sum}}^{\rm{w}}$,
$\nabla _{{\bf{F}}_l } R_{{\rm{sum}}}^{\rm{w}}$, $\nabla _{{\bf{P}}_l } R_{{\rm{sum}}}^{\rm{w}}$ in Algorithm 1 and evaluating
$\widetilde R_{\rm upp,\,sum}^w$ and $\nabla _{{\bf{P}}_l } \widetilde R_{\rm upp,\,sum}^w$ in Algorithm 2 with different antennas numbers is shown in
Table I--Table III. We observe from these tables that the computational effort for evaluating
 $\widetilde R_{\rm upp,\,sum}^w$ and $\nabla _{{\bf{P}}_l } \widetilde R_{\rm upp,\,sum}^w$ in Algorithm 2
is several orders of magnitude less than that  for evaluating
$R_{\rm{sum}}^{\rm{w}}$, $\nabla _{{\bf{F}}_l } R_{{\rm{sum}}}^{\rm{w}}$, and $\nabla _{{\bf{P}}_l } R_{{\rm{sum}}}^{\rm{w}}$
in Algorithm 1.

\begin{table}[!ht]

\label{runing_time_1}
\centering
 \captionstyle{center}
  {
\caption{The running time for evaluating $R_{\rm{sum}}^{\rm{w}}$ in  Algorithm 1 and $\widetilde R_{\rm upp,\,sum}^w$
in Algorithm 2 } \label{tab:adp_result}
\begin{tabular}{|c|c|c|}
\hline
  Case  &     $R_{\rm{sum}}^{\rm{w}}$   &   $\widetilde R_{\rm upp,\,sum}^w$          \\ \hline
  $N_t = N_r = 2$  &  0.270 s  &   0.00100006 s        \\ \hline
 $N_t = N_r = 4$    &  0.315 s    &   0.00100038 s          \\ \hline
  $N_t = N_r = 6$    &  0.461 s    &   0.00100039 s          \\ \hline
   $N_t = N_r = 8$    &  0.528 s    &   0.00100045 s          \\ \hline
\end{tabular}
}
\end{table}

\begin{table}[!ht]
\label{runing_time_2}
\centering
 \captionstyle{center}
   {
\caption{The running time for evaluating $\nabla _{{\bf{F}}_l } R_{{\rm{sum}}}^{\rm{w}}$ in  Algorithm 1 } \label{tab:adp_result}
\begin{tabular}{|c|c|c|}
\hline
  Case  &     $\nabla _{{\bf{F}}_l } R_{{\rm{sum}}}^{\rm{w}}$   &   $\mathbf{F}_l$ in (27)    \\ \hline
  $N_t = N_r = 2$  &  0.223 s  &     $\times$   \\ \hline
 $N_t = N_r = 4$    &  0.356 s    &   $\times$           \\ \hline
  $N_t = N_r = 6$    &  0.426 s    &   $\times$          \\ \hline
   $N_t = N_r = 8$    &  0.529 s    &   $\times$          \\ \hline
\end{tabular}
}
\end{table}

\begin{table}[!ht]
\label{runing_time_3}
\centering
 \captionstyle{center}
   {
\caption{The running time for evaluating $\nabla _{{\bf{P}}_l } R_{{\rm{sum}}}^{\rm{w}}$ in  Algorithm 1 and
 $\nabla _{{\bf{P}}_l } \widetilde R_{\rm upp,\,sum}^w$ in  Algorithm 2} \label{tab:adp_result}
\begin{tabular}{|c|c|c|}
\hline
  Case  &    $\nabla _{{\bf{P}}_l } R_{{\rm{sum}}}^{\rm{w}}$    &   $\nabla _{{\bf{P}}_l } \widetilde R_{\rm upp,\,sum}^w$        \\ \hline
  $N_t = N_r = 2$  &  1.250 s  &   0.00199955 s        \\ \hline
 $N_t = N_r = 4$    &  2.050 s    &   0.00200003 s          \\ \hline
  $N_t = N_r = 6$    &  2.402 s    &   0.00200006 s          \\ \hline
   $N_t = N_r = 8$    &  2.924 s    &   0.00200010 s          \\ \hline
\end{tabular}
}
\end{table}

}

\subsection{Performance of the Transmission Design}

First, we consider the doubly correlated MIMO channels, which can be modeled as
\begin{equation}\label{double_correlated}
{\bf{H}}_l  = {\bf{R}}_{r,\,l}^{1/2} {\bf{H}}_w {\bf{R}}_{t,\,l}^{1/2} , \quad l = 1,2
\end{equation}
where ${\bf{H}}_w$ is a complex random matrix with independent random entries following $\mathcal{CN}(0,1)$. The matrices ${\mathbf{R}}_{r,\,l }$
and ${\mathbf{R}}_{t,\,l}$ denote the receive and transmit correlation matrices of the $l$-th receiver channels respectively.

Here we assume the receiver correlation matrices are given
by
\begin{align}\label{Rr_1}
{\bf{R}}_{r,\,1} &  = \left[ {\begin{array}{*{20}c}
   1 & { - 0.1 - 0.05j}  \\
   { - 0.1 + 0.05j} & 1  \\
\end{array}} \right], \\
 {\bf{R}}_{r, \, 2}  & = \left[ {\begin{array}{*{20}c}
   1 & {-0.05 - 0.1j}  \\
   {-0.05 + 0.1j} & 1  \\
\end{array}} \right]
\end{align}
and compare two examples with different transmit correlation matrices as follows:

1) Example 1:
\begin{align}\label{R_t_1}
{\bf{R}}_{t,\,1} & = \left[ {\begin{array}{*{20}c}
   1 & {0.85 + 0.13j }  \\
   {0.85 - 0.13j} & 1  \\
\end{array}} \right], \\
 {\bf{R}}_{t,\,2} & = \left[ {\begin{array}{*{20}c}
   1 & {-0.8 - 0.11j}  \\
   { -0.8 + 0.11j } & 1  \\
\end{array}} \right].
\end{align}

2) Example 2:
\begin{align}\label{R_t_2}
{\bf{R}}_{t,\,1}  &= \left[ {\begin{array}{*{20}c}
   1 & {0.95 + 0.12j }  \\
   {0.95 - 0.12j } & 1  \\
\end{array}} \right], \\
 {\bf{R}}_{t,\,2} & = \left[ {\begin{array}{*{20}c}
   1 & {-0.9 + 0.09j}  \\
   {-0.9 - 0.09j} & 1  \\
\end{array}} \right].
\end{align}

Figure 1 and Figure 2 plot the sum-rate performance of Example 1 and Example 2 achieved
by different transmit designs respectively. { For Algorithm \ref{Gradient_FBC_low},
we compute  the exact LAWSR in (\ref{wsum}) with the obtained ${\widetilde{{\bf{F}}}}_l$ and  ${\bf{{\Sigma} }}_l$.  Then, we plot the exact LAWSR.}
 For comparison purpose, we  plot the sum rate performances achieved by the
``TDMA" case. { Also, to evaluate our proposed design,  we propose a no-interference upper bound of the MIMO BC with statistical CSI
as follows \cite{Bennatan}:
\begin{equation}
\begin{array}{l}
\overline R _{{\rm{upper}}}  = \sum\limits_{l = 1}^L \mu_l {E\left[ {\log_2 \det \left( {{\bf{I}}_{N_r }  } \right.} \right. }  \\
\hspace{1cm} {\left. {\left. { + \left( {N_0 {\bf{I}}_{N_r }  + {\bf{H}}_l \sum\limits_{k = l + 1}^L {{\bf{\Sigma}} _k  }}
  {{{\bf{H}}_l^H } } \right)^{ - 1}  {\bf{H}}_l {\bf{\Sigma}} _l {\bf{H}}_l^H } \right)} \right]}.
 \end{array}
\end{equation}
The upper bound $\overline R _{{\rm{upper}}}$ is denoted as ``Upper Bound" in the figures.  It is noted that for the MIMO BC with statistical CSI, the upper bound
$\overline R _{{\rm{upper}}}$ coincides with the WSR of the no-interference channel,
which is clearly the best we can expect.  Nevertheless, whether this bound is achievable is still unknown. }

From Figure 1 and Figure 2, we can make several observations.

\begin{itemize}
\item[1)] The transmit designs in Algorithm 1 and Algorithm 2 have better sum-rate performance than other design methods
throughout the entire SNR region.

\item[2)] The curves for maximizing the upper bound via Algorithm 2 and maximizing the exact LAWSR directly via  Algorithm 1  are
virtually the same, but the complexity is different.  The method for maximizing the exact  LAWSR requires, at each iteration,
numerically averaging certain random matrices involving the inverse of instantaneous realizations of the MIMO channels.
Thus, the computing effort of Algorithm 2, which does not require such numerical averaging, is significantly less than that of Algorithm 1.

\item[3)] The proposed designs offer appreciable gains in sum-rate performance when
compared against the ``TDMA" design. Specifically,
at the sum-rate $10$ b/s/Hz, the SNR gains of the proposed precoding designs over the  ``TDMA" design are approximately $4.5$ dB and $7$ dB
for Example 1 and Example 2 respectively.

\item[4)] { The proposed designs perform close to the upper bound $\overline R _{{\rm{upper}}}$
and approach the bound as the transmit correlation increases. }

\end{itemize}

{ The matrices $\mathbf{F}_l$ and $\mathbf{P}_l$ obtained by Algorithm 1 and Algorithm 2 at $\rm{SNR} = 0$ dB are given by

1) Example 1:
\begin{itemize}

\item Algorithm 1

\begin{align}\label{F_example_1_Alg1}
\mathbf{F}_1 & = 0 \\
 \mathbf{F}_2 & = \left[ {\begin{array}{*{20}c}
\ \,    0.3240 + 0.0018j & -0.3206 - 0.0463j \\
  -0.3200 + 0.0462j  & \ \ \, 0.3232 - 0.0018j \\
\end{array}} \right]
\end{align}

\begin{align}\label{P_example_1_Alg1}
{\bf{P}}_{1} & = \left[ {\begin{array}{*{20}c}
  -0.2712 + 0.3459j &   0.2300 - 0.0272j \\
  -0.2215 + 0.3845j &  0.2242 - 0.0590j \\
\end{array}} \right] \\
 {\bf{P}}_{2} & = \left[ {\begin{array}{*{20}c}
\ \,     0.3406 - 0.2130j & -0.1300 - 0.2710j \\
  -0.3051 + 0.2603j & \ \ \, 0.1687 + 0.2480j \\
\end{array}} \right]
\end{align}

\item Algorithm 2
\begin{align}\label{F_example_1_Alg2}
\mathbf{F}_1 & = 0 \\
 \mathbf{F}_2 & = \left[ {\begin{array}{*{20}c}
\ \,    0.3240 + 0.0018j & -0.3206 - 0.0463j \\
  -0.3200 + 0.0462j  & \ \ \, 0.3232 - 0.0018j \\
\end{array}} \right]
\end{align}

\begin{align}\label{P_example_1_Alg2}
{\bf{P}}_{1} & = \left[ {\begin{array}{*{20}c}
  -0.2657 + 0.3435j  & 0.2280 - 0.0289j \\
  -0.2244 + 0.3838j &  0.2241 - 0.0570j \\
\end{array}} \right] \\
 {\bf{P}}_{2} & = \left[ {\begin{array}{*{20}c}
 \ \,    0.3422 - 0.2143j & -0.1301 - 0.2727j \\
  -0.3067 + 0.2613j & \ \ \, 0.1699 + 0.2488j \\
\end{array}} \right]
\end{align}

\end{itemize}

2) Example 2:
\begin{itemize}
\item Algorithm 1

\begin{align}\label{F_example_2_Alg1}
\mathbf{F}_1 & = 0 \\
 \mathbf{F}_2 & = \left[ {\begin{array}{*{20}c}
  \ \,    0.1779 + 0.0023j & -0.1879 + 0.0347j  \\
  -0.3916 - 0.0373j & \ \ \,  0.3853 + 0.0162j  \\
\end{array}} \right]
\end{align}

\begin{align}\label{P_example_2_Alg1}
{\bf{P}}_{1} & = \left[ {\begin{array}{*{20}c}
 -0.2712 + 0.3459j &  0.2300 - 0.0272j  \\
  -0.2215 + 0.3845j  & 0.2242 - 0.0590j \\
\end{array}} \right] \\
 {\bf{P}}_{2} & = \left[ {\begin{array}{*{20}c}
\ \, 0.3406 - 0.2130j &  -0.1300 - 0.2710j \\
  -0.3051 + 0.2603j & \ \ \, 0.1687 + 0.2480j \\
\end{array}} \right]
\end{align}

\item Algorithm 2

\begin{align}\label{F_example_2_Alg2}
\mathbf{F}_1 & = 0 \\
 \mathbf{F}_2 & = \left[ {\begin{array}{*{20}c}
  \ \, 0.3240 + 0.0018j & -0.3206 - 0.0463j \\
  -0.3200 + 0.0462j & \ \ \, 0.3232 - 0.0018j \\
\end{array}} \right]
\end{align}

\begin{align}\label{P_example_2_Alg2}
{\bf{P}}_{1} & = \left[ {\begin{array}{*{20}c}
  -0.2657 + 0.3435j &  0.2280 - 0.0289j \\
  -0.2244 + 0.3838j  &  0.2241 - 0.0570j \\
\end{array}} \right] \\
 {\bf{P}}_{2} & = \left[ {\begin{array}{*{20}c}
 \ \,  0.3422 - 0.2143j & -0.1301 - 0.2727j \\
  -0.3067 + 0.2613j & \ \ \, 0.1699 + 0.2488j \\
\end{array}} \right]
\end{align}

\end{itemize}
It should be noted that $\mathbf{F}_l$ and $\mathbf{P}_l$ obtained by Algorithm 1 and Algorithm 2 maximize
the exact LAWSR $R_{\rm sum}^{\rm w}$ and the upper bound $\widetilde R_{\rm upp,\,sum}^w$, respectively.
Although $\widetilde R_{\rm upp,\,sum}^w$ is close to $\widetilde R_{\rm upp,\,sum}^w$ due to the canceling
effect of the bounding error, they are not the same. As a result, the matrices $\mathbf{F}_l$ and $\mathbf{P}_l$ obtained by Algorithm 1 and Algorithm 2 are close, but not always the same.
}

\begin{figure}[!h]
\centering
\includegraphics[width=0.5\textwidth]{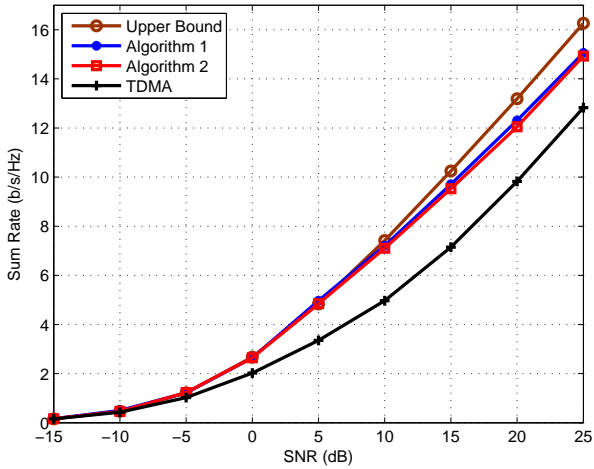}
\caption {\space\space Sum rates in the fading MIMO BC for different transmit designs and Example 1. }
\end{figure}

\begin{figure}[!h]
\centering
\includegraphics[width=0.5\textwidth]{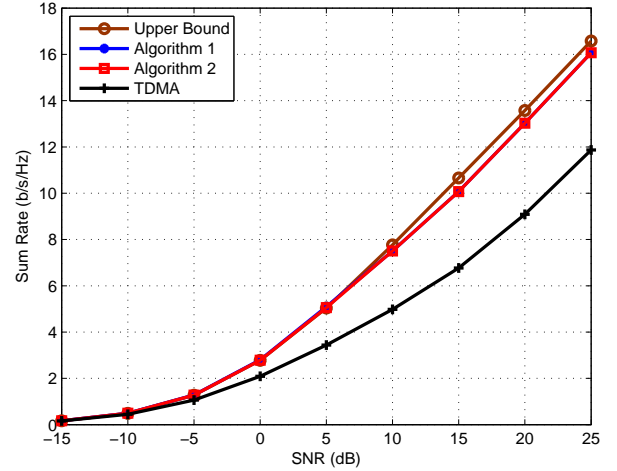}
\caption {\space\space Sum rates in the fading MIMO BC for different transmit designs and Example 2. }
\end{figure}

To further validate the proposed designs, we provide Example 3 and Example 4 as follows.
We assume the receiver correlation matrices in both examples are given  in (\ref{Rr_1}), (\ref{Rr_2})
at the top of the next page. Transmit correlation matrices of the two examples are given in (\ref{R_t_1}),
(\ref{R_t_2}) and (\ref{R_t_11}),  (\ref{R_t_22}) at the top of the next page.

\begin{figure*}[!t]
\begin{equation}\label{Rr_1}
{\bf{R}}_{r,\, 1}  = \left[ {\begin{array}{*{20}c}
  1 & {-0.12 - 0.18j} &  \ \ {0.08 + 0.05j} &  { -0.02 - 0.13j}  \\
 { -0.12 + 0.18j} & {1}  & {-0.17 - 0.16j} & \ \ {0.11 + 0.04j}  \\
  \ \ { 0.08 - 0.05j} & { -0.17 + 0.16j} &  {1} & { -0.17 - 0.16j} \\
     {  -0.02 + 0.13j}  & \ \ { 0.11 - 0.04j} &   {-0.17 + 0.16j}  & {1} \\
\end{array}} \right],
\end{equation}
\begin{equation}\label{Rr_2}
 {\bf{R}}_{r, \, 2} = \left[ {\begin{array}{*{20}c}
1 & -0.11 + 0.15j & \ \ 0.07 + 0.04j &  -0.01 - 0.10j \\
     -0.11 - 0.15j & 1  & \ \ 0.10 + 0.10j & \ \ 0.05 - 0.02j \\
  \ \     0.07 - 0.04j   & \ \ 0.10 - 0.10j  & 1 &  -0.10 - 0.20j \\
     -0.01 + 0.10j  & \ \ 0.05 + 0.02j &  -0.10 + 0.20j  & 1 \\
\end{array}} \right].
\end{equation}
 \hrulefill
\vspace*{4pt}
\end{figure*}

\begin{figure*}[!t]
\begin{equation}\label{R_t_1}
{\bf{R}}_{t,\,1}  = \left[ {\begin{array}{*{20}c}
 1       &     0.61 + 0.34j  & 0.28         &    0.61 - 0.34j \\
 0.61 - 0.34j  & 1       &     0.61 + 0.34j  &  0.28          \\
 0.28         &    0.61 - 0.34j  & 1    &        0.61 + 0.34j \\
0.61+ 0.34j  & 0.28    &        0.61 - 0.34j    & 1    \\
\end{array}} \right],
\end{equation}
\begin{equation}\label{R_t_2}
{\bf{R}}_{t,\,2}  = \left[ {\begin{array}{*{20}c}
   1       &    -0.24 - 0.71j & -0.48          &  -0.24 + 0.71j \\
  -0.24 + 0.71j &  1      &      -0.24 - 0.71j  & -0.48    \\
 -0.48        &    -0.24 + 0.71j  & 1       &    -0.24 - 0.71j \\
-0.24 - 0.71j &  -0.48       &      -0.24 + 0.71j &   1\\
\end{array}} \right].
\end{equation}
 \hrulefill
\vspace*{4pt}
\end{figure*}

\begin{figure*}[!t]
\begin{equation}\label{R_t_11}
{\bf{R}}_{t,\,1}  = \left[ {\begin{array}{*{20}c}
 1       &    { 0.94 + 0.01j } &   { 0.93 } &          {  0.94 - 0.01j} \\
{ 0.94 - 0.01j}   &    {1    } &      {   0.94 + 0.01j} & { 0.93 }  \\
 0.93     &        0.94 - 0.01j   & 1     &     0.94 + 0.01j     \\
 0.94 + 0.01j  & 0.93         &    0.94 - 0.01j   & 1  \\
\end{array}} \right],
\end{equation}
\begin{equation}\label{R_t_22}
{\bf{R}}_{t,\,2}  = \left[ {\begin{array}{*{20}c}
  1    &         0.00 - 0.92j  & -0.92         &     0.00 + 0.92j \\
  0.00 + 0.92j  & 1      &       0.00 - 0.92j  & -0.92      \\
-0.92      &      0.00 + 0.92j  & 1       &     0.00 - 0.92j \\
 0.00 - 0.92j  &  -0.92      &        0.00 + 0.92j  &  1   \\
\end{array}} \right].
\end{equation}
 \hrulefill
\vspace*{4pt}
\end{figure*}

Figure 3 and Figure 4 compare the sum-rate performance given different transmit designs.  These
results show similar observations as with Example 1 and Example 2. We find that the transmit design based on the upper bound in Proposition 1 performs nearly identically to the design based on
the exact result.  Moreover, the proposed designs outperform the ``TDMA" design throughout the entire SNR region.
To achieve a target sum-rate
of $15$ b/s/Hz,  the SNR gains of the proposed designs over the ``TDMA" design are almost $5.2$ dB and $7.5$ dB for Example 3 and Example 4 respectively.
Also, { the performance of the proposed designs is close to the upper bound $\overline R _{{\rm{upper}}}$.}

\begin{figure}[!h]
\centering
\includegraphics[width=0.5\textwidth]{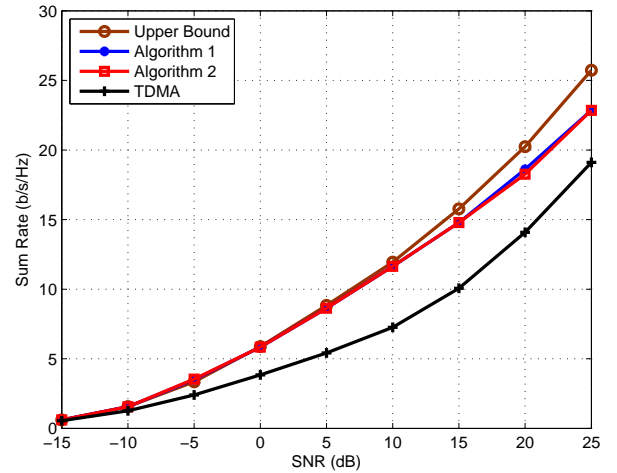}
\caption {\space\space Sum rates in the fading MIMO BC for  different transmit designs and Example 3. }
\end{figure}

\begin{figure}[!h]
\centering
\includegraphics[width=0.5\textwidth]{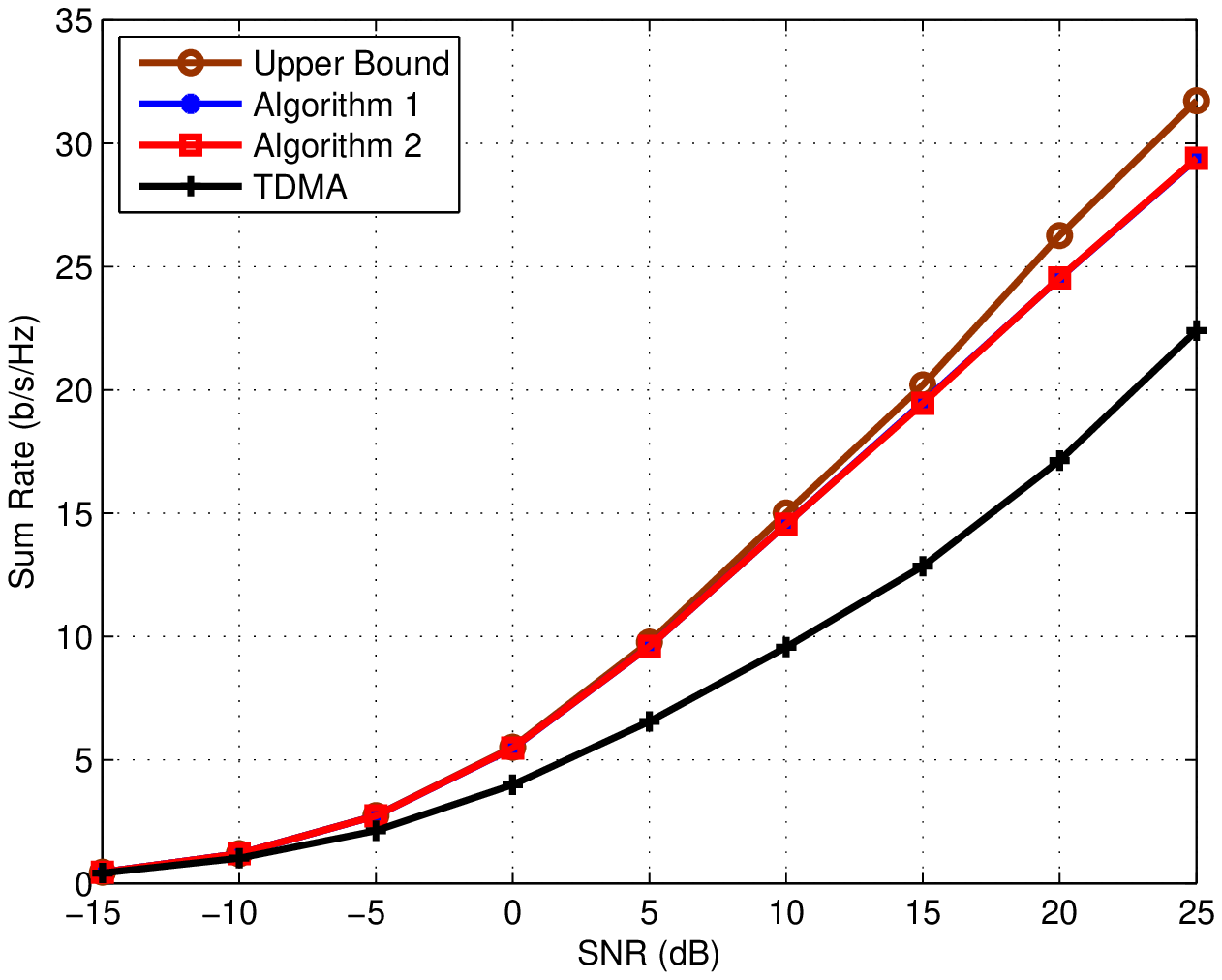}
\caption {\space\space Sum rates in the fading MIMO BC for  different transmit designs and Example 4. }
\end{figure}

\begin{figure}[!h]
\centering
\includegraphics[width=0.5\textwidth]{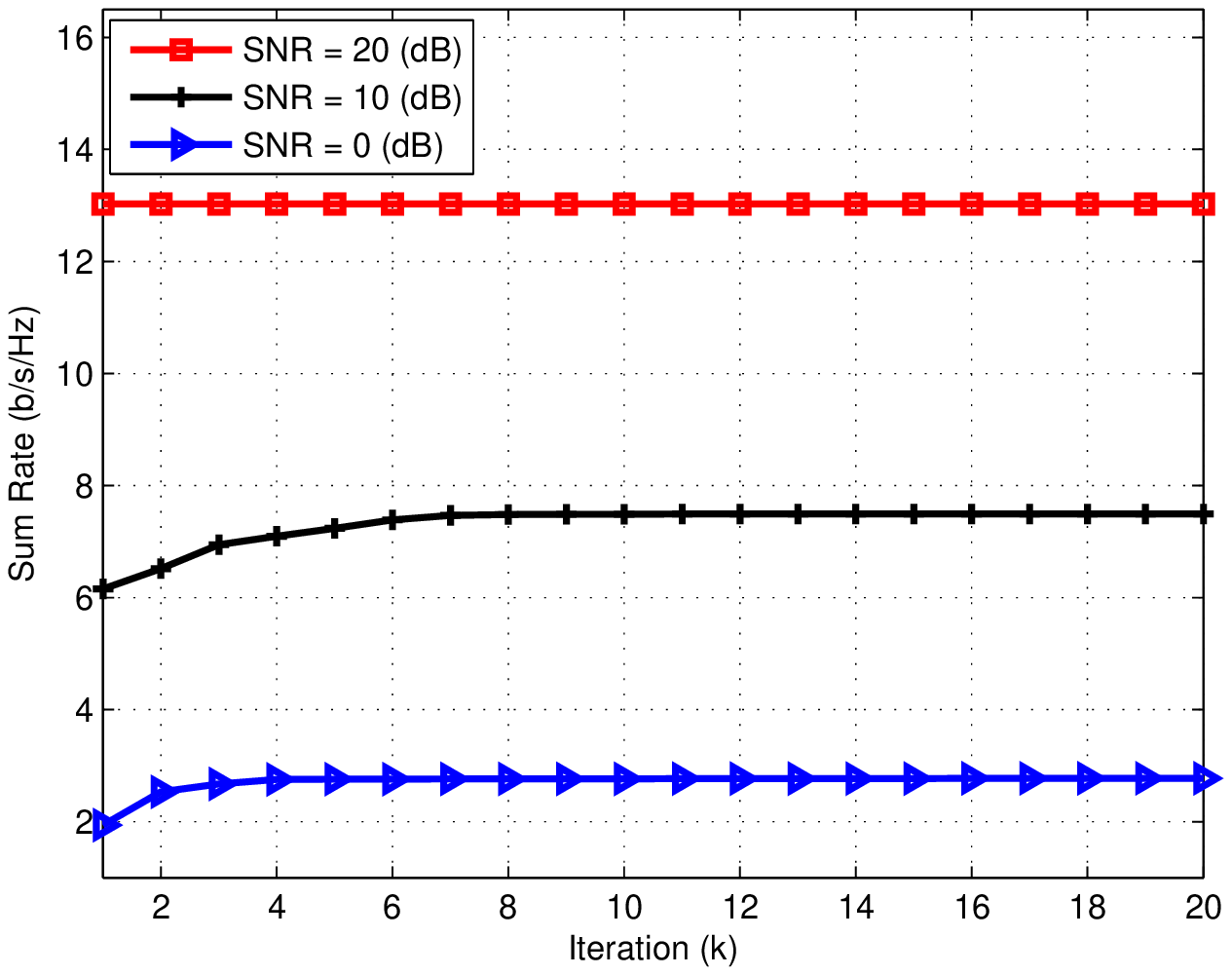}
\caption {\space\space Convergence of Algorithm 1 for Example 2 at different SNR levels.}
\end{figure}

\begin{figure}[!h]
\centering
\includegraphics[width=0.5\textwidth]{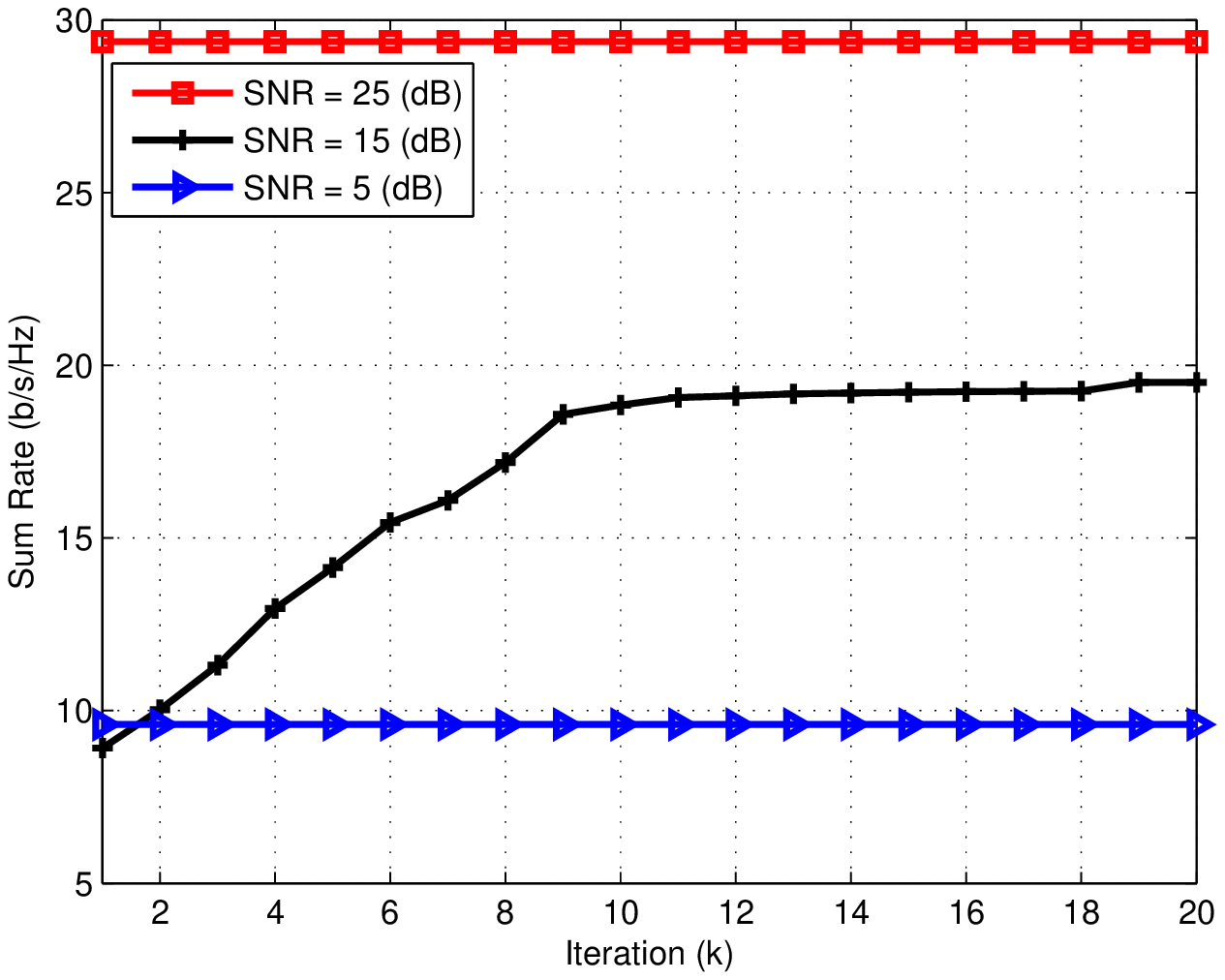}
\caption {\space\space Convergence of Algorithm 1 for Example 4 at different SNR levels.}
\end{figure}

Figure 5 and Figure 6 illustrate the convergence behavior of Algorithm 1 for different examples at different SNR levels.  We can see that in
all cases considered, Algorithm 1 converges  within a few steps. { It should be noted that for some SNR levels,
 the initialization provided by Algorithm 2 offers the maximal LAWSR. As a result,
the sum rate barely increases during the iteration for these SNR levels.  { The fast convergence behavior of Algorithm 1 implies
it can potentially be implemented in practice.  However, as indicated above, in each iteration step, Algorithm 1 requires Monte-Carlo procedure.
This will increase the implementation complexity. Henceforth, Algorithm 1 can also be utilized to provide a performance limit
criterion for other precoding designs in practical systems.}
To reduce the complexity, we further propose Algorithm 2 which does not need numerical averaging.  }

In practical scenarios, the channels often include line-of-sight (LOS) paths. To verify the LOS impact on the accuracy of the upper bound,
we consider  the Rician fading channel model
\begin{equation}\label{Rician_model}
{\bf{H}}_l  = \sqrt {\frac{K}{{K + 1}}} \overline {\bf{H}} _l  + \sqrt {\frac{1}{{K + 1}}}  {\bf{R}}_{r,\,l}^{1/2} {\bf{H}}_w {\bf{R}}_{t,\,l}^{1/2} , \quad l = 1,2
\end{equation}
where matrix ${\bf{H}}_l $ is a deterministic matrix, satisfying $\tr(\overline {\bf{H}} _l  \overline {\bf{H}} _l ^H) = N_r N_t$, and
$K \geq 0$ is the Rician $K$-factor.  We assume
\begin{equation}\label{Rician_model_2}
\overline {\bf{H}} _1  = {\rm{ }}\left[ {\begin{array}{*{20}c}
   {{\rm{0}}{\rm{.5898}}} & {{\rm{1}}{\rm{.1795}}}  \\
   {{\rm{0}}{\rm{.2949}}} & {{\rm{ 1}}{\rm{.4744}}}  \\
\end{array}} \right], \quad \overline {\bf{H}} _2  = {\rm{ }}\left[ {\begin{array}{*{20}c}
   {{\rm{0}}{\rm{.3849}}} & {{\rm{1}}{\rm{.1547}}}  \\
   {{\rm{0}}{\rm{.3849}}} & {{\rm{1}}{\rm{.5396}}}  \\
\end{array}} \right]{\rm{ }}.
\end{equation}
$\mathbf{R}_{r,\,l}$ and $\mathbf{R}_{t,\,l}$ are chosen as in Example 1. Figure 7 shows the sum rate
performance of Algorithm\footnote{The main purpose here is to examine the accuracy of the upper bound as the $K$-factor increases. Thus, we only simulate Algorithm 2 in Figure 7.} 2 in the context of different $K$-factors and SNRs. The sum rates achieved by the iterative water-filling algorithm \cite{Jindal} with instantaneous channel matrices,
where $ {\bf{H}} _l = \overline {\bf{H}} _l$, $l = 1, 2$, are also plotted as benchmarks.
We observe from Figure 7 that the sum rate performance of Algorithm 2 improves and approaches the sum rates obtained with instantaneous channel matrices as $K$ increases. This is expected because
if the channels of all  receivers converge to a constant, (\ref{R_upper_F}) indicates that the design based on Theorem \ref{upper_bound}
tends to be optimal and  Algorithm 2 becomes  equivalent to the DPC design in \cite{Jindal}.

\begin{figure}[!h]
\centering
\includegraphics[width=0.5\textwidth]{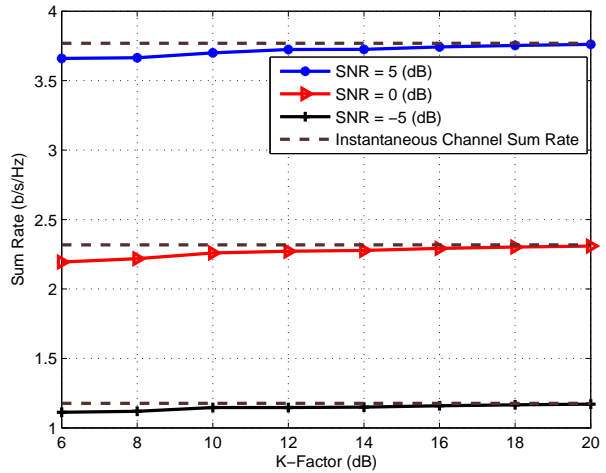}
\caption {\space\space Sum rates in the fading MIMO BC for different $K$-factors and SNRs via Algorithm 2. }
\end{figure}

\section{Conclusion}
This paper has considered the transmit design over the fading MIMO BC with statistical CSI.
To address this problem, a linear assignment operation was implemented. We first determined
a set of necessary conditions for the optimal transmit design, from which an iterative gradient descent
algorithm was developed to maximize the LAWSR but with a high computational complexity.
Thus, we employed Jensen's inequality into two substraction terms of the LAAR for each receiver to average the random component
and proved that this results in a strict upper bound of the LAAR.
In light of this, a concise closed-form expression of the linear assignment
matrix was derived for each receiver in terms of  second-order statistics of the CSI.
The derived expression captures some well known construction properties of the DPC design
in the BC model with  instantaneous CSI at the transmitter and has achieved a similar interference mitigation effect as DPC.
This immediately permits the application of classical maximum weighted sum-rate algorithms in
the MIMO BC model to design the precoding matrices. Then, we formulated a low-complexity  transmission scheme via second-order statistics of the CSI
based on the obtained precoding matrices and the closed-form linear assignment
matrices.
Moreover, transmit strategies in two special fading channel models were discussed.
Finally, we provided
concrete simulation results to illustrate the substantial gains achieved by the proposed designs.

\appendices
\section{Proof of the Theorem \ref{nec_cond}} \label{proof:nec_cond}
We first rewrite $\boldsymbol \Sigma_l = \mathbf{P}_l  \mathbf{P}_l^{H}$.
Let $\theta$ be the Lagrange multiplier associated with the inequality constraint
${\sum\nolimits_{l = 1}^L {{\rm{tr}}\left( {{\mathbf{P}}_l {\mathbf{P}}_l^H } \right) \leq P}}$.  Then, we have the cost function for
the optimal design as
\begin{equation}\label{cost_fun}
L\left({\bf{F}}, {{\bf{P}},\lambda } \right) =  - R_{{\rm{sum}}}^{\rm{w}}  +   \theta \left( {\sum\limits_{l = 1}^L {{\rm{tr}}\left( {{\bf{P}}_l {\bf{P}}_l^H } \right)}  - P} \right).
\end{equation}
By employing similar approaches as those in \cite{Palomar,Christensen}, we define the complex gradient operator as $\nabla _{{\mathbf{W}}} f = \frac{{\partial f}}{{\partial {\mathbf{W}}^* }}$. The $\left( {i,j} \right)^{\rm{th}}$
element of the matrix $\mathbf{W}$ with the complex gradient operator is defined as $\left\{ {\nabla _{{\mathbf{W}}} f} \right\}_{i,j}  = \nabla _{\left\{ {\mathbf{W}} \right\}_{i,j} } f = \frac{{\partial f}}{{\partial \left\{ {{\mathbf{W}}^* } \right\}_{i,j} }}$.  To this end, the KKT conditions satisfied by the optimal $\mathbf{F}_l$, $\mathbf{P}_l$, $l = 1,2,\cdots,L$, and $\theta$ can be expressed as { {\cite[Eq. (5.49)]{Boyd}}}
\begin{equation}\label{der_1}
\nabla _{{\bf{F}}_l } L\left( {{\bf{F}},{\bf{P}},\theta } \right) = 0, \quad l = 1,2, \cdots ,L
\end{equation}
\begin{equation}\label{der_2}
\nabla _{{\bf{P}}_t } L\left( {{\bf{F}},{\bf{P}},\theta } \right) = 0, \quad t = 1,2, \cdots ,L
\end{equation}
\begin{equation}\label{der_3}
\theta \left( {\sum\limits_{l = 1}^L {\rm{tr}} \left({{\bf{P}}_l {\bf{P}}_l^H } \right) - P} \right) = 0
\end{equation}
\begin{equation}\label{der_4}
{\sum\limits_{l = 1}^L {\rm{tr}} \left({{\bf{P}}_l {\bf{P}}_l^H } \right) - P} \leq 0
\end{equation}
\begin{equation}\label{der_5}
\theta  \ge 0.
\end{equation}

Next, we consider the calculation of $\nabla _{{\bf{F}}_l } L\left( {{\bf{F}},{\bf{P}},\theta } \right)$.  We first express the
 $(m,n)^{\rm{th}}$ element  of the matrix $\mathbf{C}_l$ as
 { \begin{equation}\label{C_F_1}
 \begin{array}{lll}
\left\{ {{\bf{C}}_l } \right\}_{m,\, n} &{ =}& {\bf{e}}_m^H \left( {{\bf{F}}_l {\bf{\Sigma }}_{S,\, l} {\bf{F}}_l^H  + {\bf{\Sigma }}_l } \right){\bf{e}}_n  \\
&{ =}& {\rm{tr}}\left( {\left( {{\bf{F}}_l {\bf{\Sigma }}_{S,\, l} {\bf{F}}_l^H   + {\bf{\Sigma }}_l } \right){\bf{e}}_n {\bf{e}}_m^H } \right) \\
&{ =}&  {\rm{tr}}\left( { {{\bf{F}}_l {\bf{\Sigma }}_{S,\, l} {\bf{F}}_l^H } {\bf{e}}_n {\bf{e}}_m^H } \right) + {\rm{tr}}\left( {\bf{\Sigma }}_l {\bf{e}}_n {\bf{e}}_m^H \right)
\end{array}
\end{equation}}
where ${\bf{e}}_m$ is a unit-vector with one at the $m^{\rm{th}}$ element and zeros elsewhere.  According to the
 complex matrix differentiation results  { \cite[Table 4.3]{Hjorungnes2011}}, we have
 {
  \begin{align}\label{C_F_2}
   & \frac{{\partial \left\{ {{\bf{C}}_l } \right\}_{m,\,n} }}{{\partial {\bf{F}}_l^* }} = \frac{{\partial {\rm{tr}}\left( { {{\bf{F}}_l {\bf{\Sigma }}_{S,\, l} {\bf{F}}_l^H } {\bf{e}}_n {\bf{e}}_m^H } \right) }}{{\partial {\bf{F}}_l^* }} = {\bf{e}}_n {\bf{e}}_m^H {\bf{F}}_l {\bf{\Sigma }}_{S,\, l} \\
\label{C_F_3}
& \frac{{\partial \left\{ {{\bf{C}}_l } \right\}_{m,\, n} }}{{\partial \left\{ {{\bf{F}}_l^* } \right\}_{i,\,j} }} = {\bf{e}}_i^H {\bf{e}}_n {\bf{e}}_m^H {\bf{F}}_l {\bf{\Sigma }}_{S,\,l} {\bf{e}}_j  = {\bf{e}}_m^H {\bf{F}}_l {\bf{\Sigma }}_{S,\,l} {\bf{e}}_j {\bf{e}}_i^H {\bf{e}}_n \\
\label{C_F_4}
& \frac{{\partial  {{\bf{C}}_l } }}{{\partial \left\{ {{\bf{F}}_l^* } \right\}_{i,\, j} }} = {\bf{F}}_l {\bf{\Sigma }}_{S,\,l} {\bf{e}}_j {\bf{e}}_i^H.
\end{align} }

Following similar steps as those in (\ref{C_F_2})--(\ref{C_F_4}), we can obtain
\begin{equation}\label{A_F_1}
\frac{{\partial \left( {{\bf{T}}_l \left( {{\bf{H}}_l } \right){\bf{A}}_l^H } \right)}}{{\partial \left\{ {{\bf{F}}_l^* } \right\}_{i,\,j} }} = {\bf{T}}_l \left( {{\bf{H}}_l } \right){\bf{\Sigma }}_{S,\,l} {\bf{e}}_j {\bf{e}}_i^H.
\end{equation}

{ {Based on the expression for the LAAR in (\ref{achievable_rate_2}), we know that $\mathbf{F}_l$ has no relation to $R_k$ for $l \neq k$.
As a result, recalling the expression for $L\left( {{\bf{F}},{\bf{P}},\theta } \right)$
in (\ref{cost_fun}) and the definition $R_{{\rm{sum}}}^{\rm{w}} = \sum\nolimits_{l = 1}^{L} \mu_l R_l$,
we have $\nabla _{{\bf{F}}_l } L\left( {{\bf{F}},{\bf{P}},\theta } \right) = \mu_l \nabla _{{\bf{F}}_l } R_l$. By exploiting
the expression for $R_l$ in (\ref{achievable_rate_2}) and the matrix derivative rule \cite[Eq. (38)]{Petersen}, it yields
(\ref{R_F_1}) and (\ref{R_F_12})  at the top of the next page.
\begin{figure*}[!t]
{
\begin{align}
\left\{ {\nabla _{{\bf{F}}_l } R_l } \right\}_{i, \, j}  &=  - \log _2 e \, E_{\mathbf{H}_l} \left[{\rm{tr}}\left( {\boldsymbol \Sigma _{{\bf{u}}_l \left| {{\bf{y}}_l , \, {\bf{H}}_l } \right.} \left( {{\bf{H}}_l } \right)^{ - 1} } \left(\frac{{\partial  {{\bf{C}}_l } }}{{\partial \left\{ {{\bf{F}}_l^* } \right\}_{i, \, j} }} - \frac{{\partial \left( {{\bf{T}}_l \left( {{\bf{H}}_l } \right){\bf{A}}_l^H } \right)}}{{\partial \left\{ {{\bf{F}}_l^* } \right\}_{i, \, j} }}  \right)\right) \right] \label{R_F_1}\\
 & = - \log _2 e \, E_{\mathbf{H}_l} \left[ {\rm{tr}} \left( {\boldsymbol \Sigma _{{\bf{u}}_l \left| {{\bf{y}}_l , \, {\bf{H}}_l } \right.} \left( {{\bf{H}}_l } \right)^{ - 1} \left[ {{\bf{F}}_l {\bf{\Sigma }}_{S,\,l} {\bf{e}}_j {\bf{e}}_i^H  - {\bf{T}}_l \left( {{\bf{H}}_l } \right){\bf{\Sigma }}_{S,\,l} {\bf{e}}_j {\bf{e}}_i^H } \right]} \right) \right]. \label{R_F_12}
\end{align}
 \hrulefill
\vspace*{4pt}
}
\end{figure*}
Then, we have
\begin{equation}\label{R_F_2}
\begin{array}{l}
\nabla _{{\bf{F}}_l } L\left( {{\bf{F}},{\bf{P}},\theta } \right)  = \mu_l \nabla _{{\bf{F}}_l } R_l = \\
 - \mu_l \log _2 e  E_{\mathbf{H}_l} \left[\left( {\boldsymbol \Sigma _{{\bf{u}}_l \left| {{\bf{y}}_l ,\, {\bf{H}}_l } \right.} \left( {{\bf{H}}_l } \right)^{ - 1} \left( {{\bf{F}}_l  - {\bf{T}}_l \left( {{\bf{H}}_l } \right)} \right){\bf{\Sigma }}_{S,\, l} } \right) \right].
\end{array}
\end{equation}

}}

Next, we  evaluate $\nabla _{{\bf{P}}_t } L\left( {{\bf{F}},{\bf{P}},\theta } \right)$.  By employing the matrix derivative rule in \cite[Eq. (33)]{Petersen}
and following similar steps as those in (\ref{C_F_2})--(\ref{R_F_2}), we have

1) $t < l$
\begin{equation}\label{R_P_1}
\begin{array}{l}
 \left\{ \! {\nabla _{{\bf{P}}_t } R_l } \! \right\}_{i, \, j} \! \! = \! \! - \log _2 e \, E \left[\! \tr\left( \! {\boldsymbol \Sigma _{{\bf{u}}_l \left| {{\bf{y}}_l ,\, {\bf{H}}_l } \right.} \! \left( {{\bf{H}}_l } \right)^{ - 1} \! {\bf{F}}_l {\bf{P}}_t {\bf{e}}_j {\bf{e}}_i^H {\bf{F}}_l^H } \!\right) \right. \\
 \hspace{1.5cm} - \tr\left({\boldsymbol \Sigma _{{\bf{u}}_l \left| {{\bf{y}}_l , \, {\bf{H}}_l } \right.} \left( {{\bf{H}}_l } \right)^{ - 1} {\bf{F}}_l {\bf{P}}_t {\bf{e}}_j {\bf{e}}_i^H {\bf{T}}_l^H \left( {{\bf{H}}_l } \!\right)}\! \right)  \\
 \hspace{0.5cm} \left.    + \tr\left( {\boldsymbol \Sigma _{{\bf{u}}_l \left| {{\bf{y}}_l , \, {\bf{H}}_l } \right.} \left( {{\bf{H}}_l } \right)^{ - 1} {\bf{T}}_l^H \left( {{\bf{H}}_l } \right){\bf{H}}_l {\bf{P}}_t {\bf{e}}_j {\bf{e}}_i^H {\bf{T}}_l \left( {{\bf{H}}_l } \right)} \right) \right] \\
 \hspace{0.5cm}  -  \tr\left( {\boldsymbol \Sigma _{{\bf{u}}_l \left| {{\bf{y}}_l ,\, {\bf{H}}_l } \right.} \left( {{\bf{H}}_l } \right)^{ - 1} {\bf{T}}_l \left( {{\bf{H}}_l } \right){\bf{H}}_l {\bf{P}}_t {\bf{e}}_j {\bf{e}}_i^H {\bf{F}}_l^H } \right)
 \end{array}
\end{equation}
\begin{equation}\label{R_P_2}
\begin{array}{l}
 \nabla _{{\bf{P}}_t } R_l  =  - \log _2 e \, E \left[{\bf{F}}_l^H  \boldsymbol\Sigma _{{\bf{u}}_l \left| {{\bf{y}}_l , \, {\bf{H}}_l } \right.} \left( {{\bf{H}}_l } \right)^{ - 1} {\bf{F}}_l {\bf{P}}_t \right.   \\
 \! - \! {\bf{T}}_l^H \left( {{\bf{H}}_l } \right) \! \boldsymbol \Sigma _{{\bf{u}}_l \left| {{\bf{y}}_l , \, {\bf{H}}_l } \right.}\! \left( {{\bf{H}}_l } \right)^{ - 1} \! {\bf{F}}_l \! {\bf{P}}_t \! \! + \! \! {\bf{T}}_l^H \left( {{\bf{H}}_l } \right)\! \boldsymbol \Sigma _{{\bf{u}}_l \left| {{\bf{y}}_l ,\,{\bf{H}}_l } \right.}\! \left( {{\bf{H}}_l } \right)^{ - 1}  \\
\hspace{1cm} \left. \times {\bf{T}}_l \left( {{\bf{H}}_l } \right){\bf{P}}_t - {\bf{F}}_l^H \boldsymbol\Sigma _{{\bf{u}}_l \left| {{\bf{y}}_l , \, {\bf{H}}_l } \right.} \left( {{\bf{H}}_l } \right)^{ - 1} {\bf{T}}_l \left( {{\bf{H}}_l } \right){\bf{P}}_t \right]. \\
 \end{array}
\end{equation}

2) $t = l$
\begin{equation}\label{R_P_3}
\begin{array}{l}
 \left\{ {\nabla _{{\bf{P}}_t } R_l } \right\}_{i, \, j}  = \log _2 e \, \tr\left( {\left( {{\bf{P}}_t^H } \right)^{ - 1} {\bf{e}}_j {\bf{e}}_i^H } \right)\\
 \hspace{1cm} -  \log _2 e \, E\left[\tr\left( {\boldsymbol \Sigma _{{\bf{u}}_l \left| {{\bf{y}}_l , \, {\bf{H}}_l } \right.} \left( {{\bf{H}}_l } \right)^{ - 1} {\bf{P}}_t {\bf{e}}_j {\bf{e}}_i^H } \right) \right. \\
 \hspace{1cm}    -\tr\left( {\boldsymbol\Sigma _{{\bf{u}}_l \left| {{\bf{y}}_l , \, {\bf{H}}_l } \right.} \left( {{\bf{H}}_l } \right)^{ - 1} {\bf{P}}_t {\bf{e}}_j {\bf{e}}_i^H {\bf{T}}_l^H \left( {{\bf{H}}_l } \right)} \right) \\
  \hspace{0.5cm}   + \tr\left( {\boldsymbol\Sigma _{{\bf{u}}_l \left| {{\bf{y}}_l , \, {\bf{H}}_l } \right.} \left( {{\bf{H}}_l } \right)^{ - 1} {\bf{T}}_l \left( {{\bf{H}}_l } \right) {\bf{P}}_t {\bf{e}}_j {\bf{e}}_i^H {\bf{T}}_l^H \left( {{\bf{H}}_l } \right)} \right) \\
  \hspace{1cm} \left. - \tr\left( {\boldsymbol\Sigma _{{\bf{u}}_l \left| {{\bf{y}}_l , \, {\bf{H}}_l } \right.} \left( {{\bf{H}}_l } \right)^{ - 1} {\bf{T}}_l \left( {{\bf{H}}_l } \right){\bf{P}}_t {\bf{e}}_j {\bf{e}}_i^H } \right)\right] \\
 \end{array}
\end{equation}
\begin{equation}\label{R_P_4}
\begin{array}{l}
 \nabla _{{\bf{P}}_t } R_l  = \log _2 e  \, \left( {{\bf{P}}_t^H } \right)^{ - 1}  -\log _2 e \, E \left[\boldsymbol\Sigma _{{\bf{u}}_l \left| {{\bf{y}}_l ,\, {\bf{H}}_l } \right.} \left( {{\bf{H}}_l } \right)^{ - 1} {\bf{P}}_t   \right.  \\
\!-\! {\bf{T}}_l^H \left( {{\bf{H}}_l } \right)\boldsymbol\Sigma _{{\bf{u}}_l \left| {{\bf{y}}_l , \, {\bf{H}}_l } \right.} \! \left( {{\bf{H}}_l } \right)^{ - 1} {\bf{P}}_t \! + \! {\bf{T}}_l^H \left( {{\bf{H}}_l } \right) \boldsymbol\Sigma _{{\bf{u}}_l \left| {{\bf{y}}_l , \,{\bf{H}}_l } \right.} \! \left( {{\bf{H}}_l } \right)^{ - 1} \\
\hspace{2cm} \left. {\bf{T}}_l \left( {{\bf{H}}_l } \right){\bf{P}}_t  - \boldsymbol\Sigma _{{\bf{u}}_l \left| {{\bf{y}}_l , \, {\bf{H}}_l } \right.} \left( {{\bf{H}}_l } \right)^{ - 1} {\bf{T}}_l \left( {{\bf{H}}_l } \right){\bf{P}}_t \right]. \\
 \end{array}
\end{equation}

3) $t > l$
\begin{equation}\label{R_P_5}
\begin{array}{l}
\left\{ {\nabla _{{\bf{P}}_t } R_l } \right\}_{i, \, j}  =  - \log _2 e \, E\left[\tr\left( {\boldsymbol\Sigma _{{\bf{u}}_l \left| {{\bf{y}}_l , \, {\bf{H}}_l } \right.} \left( {{\bf{H}}_l } \right)^{ - 1} {\bf{T}}_l \left( {{\bf{H}}_l } \right){\bf{P}}_t } \right. \right. \\
\hspace{5cm} \left. \left.{{\times \bf{e}}_j {\bf{e}}_i^H {\bf{T}}_l^H \left( {{\bf{H}}_l } \right)} \right) \right]
 \end{array}
\end{equation}
\begin{equation}\label{R_P_6}
\left\{ {\nabla _{{\bf{P}}_i } R_l } \right\}  \! = \!  - \log _2 e \, E\left[{\bf{T}}_l^H \left( {{\bf{H}}_l } \right)\boldsymbol\Sigma _{{\bf{u}}_l \left| {{\bf{y}}_l , \, {\bf{H}}_l } \right.} \! \left( {{\bf{H}}_l } \right)^{ - 1} {\bf{T}}_l \left( {{\bf{H}}_l } \right) \right].
\end{equation}

Based on (\ref{wsum}) and $\nabla _{{\bf{P}}_t } \left( {\sum\nolimits_{l = 1}^L {{\bf{P}}_l {\bf{P}}_l^H } } \right) = {\bf{P}}_t$, $\nabla _{{\bf{P}}_t } L\left( {{\bf{F}},{\bf{P}},\theta } \right)$ can be expressed as
{ {
\begin{equation}\label{R_P_7}
\begin{array}{l}
\nabla _{{\bf{P}}_t } L\left( {{\bf{F}},{\bf{P}},\theta } \right) \! = \!  - \left( {\sum\limits_{l = 1}^{t - 1} \mu_l {\nabla _{{\bf{P}}_t } R_l }  + \mu_t\nabla _{{\bf{P}}_t } R_t }\right.
 \\ \hspace{2cm}  \left. {+ \sum\limits_{l = t + 1}^L {\mu_l \nabla _{{\bf{P}}_t } R_l } } \right) + \theta {\bf{P}}_t.
\end{array}
\end{equation}
} }
Finally, the theorem can be proved by combining  (\ref{der_1})--(\ref{der_5}), (\ref{R_F_2}),  (\ref{R_P_2}), (\ref{R_P_4}), (\ref{R_P_6}), and (\ref{R_P_7}), along with some simplifications.

\section{Proof of the Theorem \ref{upper_bound}} \label{proof:upper_bound}
Before we present the proof, we find the following two lemmas useful.
\begin{lemma}
Assume $\mathbf{X} \succ \mathbf{0}$ is a  $N \times N$ random matrix with $E[\mathbf{X}] = \mathbf{M}_x$.
Then, the following result holds
\begin{equation}\label{Inequality_1}
E\left[ {{\bf{X}}^{ - 1} } \right] \succeq \mathbf{M}_x^{-1}.
\end{equation}
\begin{proof}
Let $\mathcal{H}^{N}(J)$ denote the space of all Hermitian $N \times N$ matrices whose eigenvalues all fall within $J$. Then, we define a matrix function $f: J \mapsto J$, given by
\begin{equation}\label{matrix_fun}
f\left( {\bf{T}} \right) = {\bf{T}}^{ - 1}.
\end{equation}
According to \cite[Lemma 2.5]{Brinkhuis2005}, we know that $f$ is a strictly matrix convex function. Let
{ $\mathbf{T} = \mathbf{T}_1 + a (\mathbf{T}_2 - \mathbf{T}_1) = (1 - a)  \mathbf{T}_1 + a \mathbf{T}_2$}, $0 \leq a \leq 1$. Then, recalling
the definition of matrix convex function in \cite[Definition 2.2]{Brinkhuis2005}, it yields
{
\begin{equation}\label{matrix_ineq_add}
(1 - a) f(\mathbf{T}_1) + a f(\mathbf{T}_2) \succeq f({\bf{T}}).
\end{equation}
Rewriting (\ref{matrix_ineq_add}), we have }
\begin{equation}\label{matrix_ineq}
f\left( {{\bf{T}}_2 } \right) - f\left( {\bf{T}} \right) \succeq \frac{{1 - a}}{a}\left( {f\left( {\bf{T}} \right) - f\left( {{\bf{T}}_1 } \right)} \right).
\end{equation}

Considering the Taylor expansion of the matrix function in \cite[Definition 3.1]{Brinkhuis2005}, ${f\left( {\bf{T}} \right) - f\left( {{\bf{T}}_1 } \right)}$
can be expressed as
\begin{equation}\label{matrix_ineq_21}
\begin{array}{l}
 f\left( {\bf{T}} \right) - f\left( {{\bf{T}}_1 } \right) \\
  = f^{\left( 1 \right)} \left( {{\bf{T}}_1 } \right)[{\bf{T}} - {\bf{T}}_1 ] + o\left( {\left\| {{\bf{T}} - {\bf{T}}_1 } \right\|_F } \right){\bf{I}}_N  \\
  = af^{\left( 1 \right)} \left( {{\bf{T}}_1 } \right)[{\bf{T}}_2  - {\bf{T}}_1 ] + o\left( {a\left\| {\left( {{\bf{T}}_2  - {\bf{T}}_1 } \right)} \right\|_F } \right){\bf{I}}_N
  \end{array}
\end{equation}
where $f^{\left( 1 \right)} \left( {{\bf{T}}_1 }\right)$ is a Hermitian symmetric multi-linear mapping  on the space $\mathcal{H}^{N}$.  Plugging
(\ref{matrix_ineq_21}) into (\ref{matrix_ineq}) and taking the limit as $a \rightarrow 0$, the right term of (\ref{matrix_ineq}) becomes
\begin{equation}\label{matrix_ineq_31}
\begin{array}{l}
 \mathop {\lim }\limits_{a \to 0} \frac{{1 - a}}{a}\left( {f\left( {\bf{T}} \right) - f\left( {{\bf{T}}_1 } \right)} \right) \\
    =  f^{\left( 1 \right)} \left( {{\bf{T}}_1 } \right)\left[ {{\bf{T}}_2  - {\bf{T}}_1 } \right] + \mathop {\lim }\limits_{a \to 0} \frac{{\left( {1 - a} \right)}}{a}o\left( {a\left\| {\left( {{\bf{T}}_2  - {\bf{T}}_1 } \right)} \right\|_F } \right){\bf{I}}_N  \\
    =   f^{\left( 1 \right)} \left( {{\bf{T}}_1 } \right)\left[ {{\bf{T}}_2  - {\bf{T}}_1 } \right].
   \end{array}
\end{equation}
Synchronously, when $a \rightarrow 0$, the left term of (\ref{matrix_ineq}) can be written as
\begin{equation}\label{matrix_ineq_4}
\mathop {\lim }\limits_{a \to 0} f\left( {{\bf{T}}_2 } \right) - f\left( {\bf{T}} \right) = f\left( {{\bf{T}}_2 } \right) - f\left( {{\bf{T}}_1 } \right).
\end{equation}
Eqs. (\ref{matrix_ineq_31}) and (\ref{matrix_ineq_4}) imply the following result for arbitrary $N \times N$ Hermitian matrices
 ${\bf{T}}_1$ and ${\bf{T}}_2$
\begin{equation}\label{matrix_ineq_5}
 f\left( {{\bf{T}}_2 } \right) - f\left( {{\bf{T}}_1 } \right) \succeq f^{\left( 1 \right)} \left( {{\bf{T}}_1 } \right)\left[ {{\bf{T}}_2  - {\bf{T}}_1 } \right].
\end{equation}
By setting $\mathbf{T}_1 = \mathbf{M}_x$, $\mathbf{T}_2 = \mathbf{X}$, and evaluating the expectations of both sides of (\ref{matrix_ineq_5}), it yields
\begin{equation}\label{matrix_ineq_6}
E\left[ f\left( {{\bf{X}} } \right) \right] -  f\left(  \mathbf{M}_x \right) \succeq f^{\left( 1 \right)} \left( {\mathbf{M}_x} \right) \left[ E\left[{\bf{X}}\right]  - \mathbf{M}_x \right] = 0
\end{equation}
which completes the proof of the lemma.
\end{proof}
\end{lemma}

\begin{lemma}
Assume $\mathbf{Y} \succeq \mathbf{0} $ is a $N \times N$ random matrix with $E[\mathbf{Y}] = \mathbf{M}_y$. $\mathbf{A}$
and $\mathbf{B}$ are  $N \times N$ constant matrices satisfying $\mathbf{A} \succ \mathbf{0}$, $\mathbf{B} \succ \mathbf{0}$,
and $\mathbf{A} - \mathbf{B} \succeq \mathbf{0}$. Then, the following inequality holds
\begin{equation}\label{Inequality_2}
\begin{array}{l}
E\left[ {\log_2 \det \left( {{\bf{YA}} + {\bf{I}}_N} \right)} \right] - E\left[ {\log_2 \det \left( {{\bf{YB}} + {\bf{I}}_N} \right)} \right] \le \\
\hspace{1cm} \log _2 \det \left( {\mathbf{M}_y {\bf{A}} + {\bf{I}}_N} \right) - \log_2 \det \left( {\mathbf{M}_y {\bf{B}}+ {\bf{I}}_N} \right).
   \end{array}
\end{equation}
\begin{proof}
We define the function $g(\mathbf{T})$ as follows
\begin{equation}\label{Ineq_fun}
g(\mathbf{T}) = \log_2 \det \left(  \mathbf{M}_y \mathbf{T} +  \mathbf{I}_N \right) - {{E}} \left[ \log_2 \det ( \mathbf{YT}  + \mathbf{I}_N )\right].
\end{equation}
Also, we construct a composite function $h(\omega)$ as
\begin{equation}\label{Ineq_fun_2}
h\left( \omega  \right):  = g\left[{\left( {1 - \omega } \right) {\bf{B}} + \omega {\bf{A}}} \right] = g \left( \mathbf{U} \right),\quad  0 \leq \omega \leq 1.
\end{equation}
Since $ \mathbf{A} \succ \mathbf{0}$ and $\mathbf{B} \succ \mathbf{0}$, it can be readily identified that $\mathbf{U}\succ \mathbf{0}$. Moreover, from (\ref{Ineq_fun_2}), we have
\begin{equation}\label{Ineq_fun_3}
h\left( 0 \right) = g\left( {\bf{B}} \right), \quad h\left( 1 \right) = g\left( {\bf{A}} \right).
\end{equation}
Results in (\ref{Ineq_fun_3}) suggest to consider
\begin{equation}\label{Ineq_fun_4}
g\left( {\bf{A}} \right) - g\left( {\bf{B}} \right) = h\left( 1 \right) - h\left( 0 \right) = \int_0^1 {\frac{{dh\left( \omega  \right)}}{{d\omega }}} d\omega.
\end{equation}
Application of the chain rule of the derivative leads to
\begin{align}\label{Ineq_fun_51}
 \frac{{dh\left( \omega  \right)}}{{d\omega }} & = \sum\limits_{r = 1}^N {\sum\limits_{s = 1}^N {\left\{ {\frac{{\partial g\left( {\bf{U}} \right)}}{{\partial {\bf{U}}}}} \right\}_{r,s} } }  \! \! \times \! \!\left\{ {\frac{d}{{d\omega }}\left( {\omega {\bf{A}} \! \! + \! \! \left( {1 - \omega } \right){\bf{B}}} \right)} \right\}_{r, s}  \\
 \label{Ineq_fun_52}
& = \sum\limits_{r = 1}^N {\sum\limits_{s = 1}^N {\left\{ {\frac{{\partial g\left( {\bf{U}} \right)}}{{\partial {\bf{U}}}}} \right\}_{r,s} } }  \times \left\{ {{\bf{A}} - {\bf{B}}} \right\}_{r,s}.
\end{align}
Following similar approaches as those in (\ref{C_F_1})--(\ref{R_F_2}) and keeping in mind that $ \mathbf{U}\succ \mathbf{0}$, we can obtain
{
\begin{align}\label{Ineq_fun_61}
 \frac{{\partial g\left( {\bf{U}} \right)}}{{\partial {\bf{U}}}} &=  \left( {\mathbf{M}_y {\bf{U}}  + {\bf{I}}_N } \right)^{ - 1} {\bf{M}}_y  - E\left[ {\left( {{\bf{YU}} + {\bf{I}}_N } \right)^{ - 1} } {\bf{Y}} \right] \\
 \label{Ineq_fun_62}
  &= \left( {\mathbf{M}_y {\bf{U}}   + {\bf{I}}_N } \right)^{ - 1}  \left[ {\mathbf{M}_y {\bf{U}}   +  {\bf{I}}_N  - {\bf{I}}_N } \right] {\bf{U}}^{ - 1} \nonumber \\
   & \hspace{0.2cm} - E\left[ \left( {\mathbf{Y} {\bf{U}}   + {\bf{I}}_N } \right)^{ - 1}  \left[ {\mathbf{Y} {\bf{U}}   + {\bf{I}}_N  - {\bf{I}}_N } \right] {\bf{U}}^{ - 1} \right] \\
  \label{Ineq_fun_63}
 & =  {\bf{U}}^{ - 1} -  \left( {\mathbf{M}_y {\bf{U}}   + {\bf{I}}_N } \right)^{ - 1} {\bf{U}}^{ - 1} \nonumber \\
   & \hspace{1cm}  -  {\bf{U}}^{ - 1}  + E\left[ \left( {\mathbf{Y} {\bf{U}}   + {\bf{I}}_N } \right)^{ - 1}  {\bf{U}}^{ - 1} \right]\\
  & = E\left[ {\left( {{\bf{UYU}} + {\bf{U}}} \right)^{ - 1} } \right] - \left( {{\bf{UM}}_y {\bf{U}} + {\bf{U}}} \right)^{ - 1} .
\end{align}
}

Applying Lemma 1 to (\ref{Ineq_fun_63}), yields
\begin{equation}\label{Ineq_fun_7}
\frac{{\partial g\left( {\bf{U}} \right)}}{{\partial {\bf{U}}}} \succeq \mathbf{0}.
\end{equation}
Recalling $\mathbf{A} - \mathbf{B} \succeq \mathbf{0}$, the result in (\ref{Ineq_fun_7}), and the Schur Product Theorem in \cite[Theorem 7.5.3]{Horn1999}, we have
\begin{equation}\label{Ineq_fun_8}
\frac{{\partial g\left( {\bf{U}} \right)}}{{\partial {\bf{U}}}} \circ (\mathbf{A} - \mathbf{B}) \succeq \mathbf{0}
\end{equation}
which implies that
\begin{equation}\label{Ineq_fun_9}
\sum\limits_{r = 1}^N {\sum\limits_{s = 1}^N {\left\{ {\frac{{\partial g\left( {\bf{U}} \right)}}{{\partial {\bf{U}}}}} \right\}_{r,s} } }  \times \left\{ {{\bf{A}} - {\bf{B}}} \right\}_{r,s} \geq 0.
\end{equation}
Then, combining (\ref{Ineq_fun_4}), (\ref{Ineq_fun_52}), and (\ref{Ineq_fun_9}), we have
\begin{equation}\label{Ineq_fun_10}
g\left( {\bf{A}} \right) - g\left( {\bf{B}} \right) \geq 0
\end{equation}
which completes the proof of the lemma.
\end{proof}
\end{lemma}

Now we begin to prove Lemma \ref{upper_bound}.
As mentioned in Section \ref{sec:system_model}, when discussing the linear assignment capacity, similar to the assumption in \cite{Bennatan},
we assume  $\mathbf{\Sigma}_l \succ \mathbf{0}$. Henceforth, we know that $\mathbf{C}_l$ is invertible. To this end,
${\log_2 \det \left( {\bf{\Sigma}} _{{\mathbf{u}_l}\left| {{\mathbf{y}_l},{\mathbf{H}_l}} \right.} \left( {\mathbf{H}_l} \right)\right)}$ in (\ref{achievable_rate_2}) can be reexpressed as
\begin{align}\label{Prop_proof_1}
& \log_2 \det \left( {{\bf{\Sigma}} _{{\bf{u}}_l\left| {{\bf{y}}_l,{\bf{H}}_l} \right.} \left( {\bf{H}}_l \right)} \right) \nonumber \\
 &= \log_2  \det \left( {\bf{C}}_l \right) \nonumber \\
  & + \log_2 \det \left( {{\bf{I}}_{N_t} - {\bf{C}}_l^{ - 1} {\bf{A}}_l \mathbf{H}_l^H \left[ {{\bf{H}}\mathbf{B}_l\mathbf{H}_l^H  + {\bf{\Sigma}} _{Z,\,l}  } \right]^{ - 1} {\bf{H}}_l \mathbf{A}_l^H  } \right) \\
& \mathop  = \limits^{\left( a \right)}\log_2  \det \left( {\bf{C}}_l \right) \nonumber \\
& + \log_2 \det \left( {{\bf{I}}_{N_t} - {\bf{H}}_l \mathbf{A}_l^H {\bf{C}}_l^{ - 1} {\bf{A}}_l \mathbf{H}_l^H \left[ {{\bf{H}}\mathbf{B}_l\mathbf{H}_l^H   + {\bf{\Sigma}} _{Z,\,l}  } \right]^{ - 1}   } \right) \\
& =  \log_2   \det \left( {\bf{C}}_l \right)
\! + \!\log_2   \det \left( {{\bf{H}}_l \left( {{\bf{B}}_l \!\!-\!\! {\bf{A}}_l^H {\bf{C}}_l^{ - 1} {\bf{A}}}_l \right){\bf{H}}_l^H  \!\!+ \!\!{\bf{\Sigma}} _{Z,\,l} } \right) \nonumber \\
& \hspace{2cm} - \log_2 {\det \left( {{\bf{H}}_l \mathbf{B}_l \mathbf{H}_l^H  + {\bf{\Sigma}} _{Z, \, l} } \right)} \\
&= \log_2  \det \left( {\bf{C}}_l \right)  \nonumber \\
& \hspace{0.2cm} + \log_2 \det \left( {{\bf{H}}_l\left(\mathbf{D}_l +
\sum\limits_{t = l + 1}^L {\mathbf{\Sigma} _t }\right) \mathbf{H}_l^H  +  N_0 \mathbf{I}_{N_r}  } \right)  \nonumber
\end{align}
\begin{align}
& - \log_2 \det \left( {{\bf{H}}_l\left(\mathbf{B}_l +
\sum\limits_{t = l + 1}^L {\mathbf{\Sigma} _t }\right) \mathbf{H}_l^H  +  N_0 \mathbf{I}_{N_r}  } \right)  \\
& \mathop  = \limits^{\left( b \right)} \log_2  \det \left( {\bf{C}}_l \right) \nonumber  \\
& \hspace{0.2cm} + \log_2 \det \left( \mathbf{H}_l^H {{\bf{H}}_l\left(\mathbf{D}_l +
\sum\limits_{t = l + 1}^L {\mathbf{\Sigma} _t }\right)  +  N_0 \mathbf{I}_{N_t}  } \right) \nonumber \\
& \hspace{2cm} + (N_r - N_t) \log_2 N_0 \nonumber  \\
& \hspace{0.2cm} -  \log_2 \det \left(  \mathbf{H}_l^H {{\bf{H}}_l\left(\mathbf{B}_l +
\sum\limits_{t = l + 1}^L {\mathbf{\Sigma} _t }\right)  +  N_0 \mathbf{I}_{N_t}  } \right) \nonumber \\
& \hspace{2cm} - (N_r - N_t) \log_2 N_0  \label{Prop_proof_5}
\end{align}
where equalities (a) and (b) are obtained according to the determinant identity $\det(\mathbf{X} + \mathbf{AB}) = \det(\mathbf{X}) \det(\mathbf{I} + \mathbf{BX}^{-1}\mathbf{A})$.

Since $\mathbf{\Sigma}_l \succ \mathbf{0}$, it yields $ \mathbf{B}_l \succ \mathbf{0}$, $\sum\nolimits_{t = l + 1}^L {\mathbf{\Sigma} _t } \succ \mathbf{0}$, and $\mathbf{B}_l - \mathbf{D}_l = \mathbf{A}_l^H \mathbf{C}_l \mathbf{A}_l \succeq \mathbf{0} $.  Next, we need to prove
that $\mathbf{D}_l \succ \mathbf{0}$.

Let $\mathbf{u}$ and $\mathbf{v}$ be two $N \times 1$ constant non-zero vectors. We begin by considering
\begin{align}
 & \left[ {{\bf{u}}^H \quad {\bf{v}}^H } \right] \left[ {\begin{array}{*{20}c}
   {{\bf{\Sigma }}_{S,\,l} } & {{\bf{\Sigma }}_{S,\,l} {\bf{F}}_l^H }   \\
   {{\bf{F}}_l {\bf{\Sigma }}_{S,\,l} } & {{\bf{F}}_l {\bf{\Sigma }}_{S,\, l} {\bf{F}}_l^H }
\end{array}} \right]\left[ \begin{array}{l}
 {\bf{u}} \\
 {\bf{v}} \\
 \end{array} \right] \nonumber \\
&  = {\bf{u}}^H {\bf{\Sigma }}_{S,\,l} {\bf{u}} \! \!+ \!\! {\bf{u}}^H {\bf{\Sigma }}_{S,\,l} {\bf{F}}_l^H {\bf{v}} \!\!+\!\! {\bf{v}}^H {\bf{F}}_l {\bf{\Sigma }}_{S,\,l} {\bf{u}} \!+ \! {\bf{v}}^H {\bf{F}}_l {\bf{\Sigma }}_{S,\, l} {\bf{F}}_l^H {\bf{v}} \\
&  = {\bf{u}}^H \left( {{\bf{\Sigma }}_{S,\,l} {\bf{u}} + {\bf{\Sigma }}_{S,\,l} {\bf{F}}_l^H {\bf{v}}} \right) + {\bf{v}}^H {\bf{F}}_l \left( {{\bf{\Sigma }}_{S,\,l} {\bf{u}} + {\bf{\Sigma }}_{S,\,l} {\bf{F}}_l^H {\bf{v}}} \right) \\
& = \left( {{\bf{u}}^H  + {\bf{v}}^H {\bf{F}}_l } \right){\bf{\Sigma }}_{S,\, l} \left( {{\bf{u}} + {\bf{F}}_l^H {\bf{v}}} \right)  \mathop  \geq \limits^{(a)} 0  \label{Prop_proof_6}
  \end{align}
where (a) is obtained based on the fact ${\bf{\Sigma }}_{S,\, l} = \sum\nolimits_{t = 1}^{l - 1} {\bf{\Sigma }}_{t} \succ \mathbf{0}$. (\ref{Prop_proof_6}) indicates that the following
result holds
\begin{equation}\label{Prop_proof_9}
\left[ {\begin{array}{*{20}c}
   {{\bf{\Sigma }}_{S,\,l} } & {{\bf{\Sigma }}_{S,\,l} {\bf{F}}_l^H }  \\
   {{\bf{F}}_l {\bf{\Sigma }}_{S,\,l} } & {{\bf{F}}_l {\bf{\Sigma }}_{S,\,l} {\bf{F}}_l^H }  \\
\end{array}} \right] + \left[ {\begin{array}{*{20}c}
   {{\bf{\Sigma }}_l } & {{\bf{\Sigma }}_l }  \\
   {{\bf{\Sigma }}_l } & {{\bf{\Sigma }}_l }  \\
\end{array}} \right] \! \! = \! \! \left[ {\begin{array}{*{20}c}
   {{\bf{B}}_l } & {{\bf{A}}_l^H }  \\
   {{\bf{A}}_l } & {{\bf{C}}_l }  \\
\end{array}} \right] \! \succ \! \mathbf{0}.
  \end{equation}
Then, we have
\begin{equation}\label{Prop_proof_10}
\begin{array}{l}
\left[ {\begin{array}{*{20}c}
   {{\bf{I}}_N } & { - {\bf{A}}_l^H {\bf{C}}_l^{ - 1} }  \\
   \mathbf{0} & {{\bf{I}}_N }  \\
\end{array}} \right]\left[ {\begin{array}{*{20}c}
   {{\bf{B}}_l } & {{\bf{A}}_l^H }  \\
   {{\bf{A}}_l } & {{\bf{C}}_l }  \\
\end{array}} \right]\left[ {\begin{array}{*{20}c}
   {{\bf{I}}_N } & \mathbf{0}  \\
   { - {\bf{C}}_l^{ - 1} {\bf{A}}_l^H } & {{\bf{I}}_N }  \\
\end{array}} \right] \\
\qquad  = \left[ {\begin{array}{*{20}c}
   {{\bf{B}}_l  - {\bf{A}}_l^H {\bf{C}}_l {\bf{A}}_l } & \mathbf{0}  \\
   \mathbf{0} & {{\bf{C}}_l }  \\
\end{array}} \right] = \left[ {\begin{array}{*{20}c}
   {{\bf{D}}_l } & \mathbf{0}  \\
   \mathbf{0} & {{\bf{C}}_l }  \\
\end{array}} \right] \succ \mathbf{0} .
\end{array}
  \end{equation}
From (\ref{Prop_proof_10}), we know that ${{\bf{D}}_l} \succ \mathbf{0}$.

Finally, utilizing the ${\log_2 \det \left( {\bf{\Sigma}} _{{\mathbf{u}_l}\left| {{\mathbf{y}_l},{\mathbf{H}_l}} \right.} \left( {\mathbf{H}_l} \right)\right)}$ expression
in (\ref{Prop_proof_5}) and applying Lemma 2 to (\ref{achievable_rate_2}), along with some simplifications, we obtain the upper bound in Lemma \ref{upper_bound}.

\section{Proof of the Proposition \ref{upper_bound_F}} \label{proof:upper_bound_F}
First, we rewrite the upper bound expression in (\ref{bound_prop}) as given in (\ref{achievable_rate_F1})--(\ref{achievable_rate_F3})
at the top of the next page, where
\begin{figure*}[!t]
\begin{equation}\label{achievable_rate_F1}
\begin{array}{l}
 R_{{\rm{upp}},\, l}  = \log _2 \det \left( {{\bf{\Sigma }}_l } \right) - \log _2 \det \left( {{\bf{C}}_l } \right) \\
 \qquad \qquad \qquad - \log _2 \det \left( {{\bf{I}}_{N_t }  - \left( {{\bf{R}}_{g, \, l} \left( {{\bf{B}}_l  + \sum\limits_{t = l + 1}^L {{\bf{\Sigma }}_t } } \right) + N_0 {\bf{I}}_{N_t } } \right)^{ - 1} {\bf{R}}_{g,\, l} {\bf{A}}_l^H {\bf{C}}_l^{ - 1} {\bf{A}}_l } \right) \\
 \end{array}
 \end{equation}
\begin{equation}\label{achievable_rate_F2}
\begin{array}{l}
\qquad \qquad \ \  = \log _2 \det \left( {{\bf{\Sigma }}_l } \right) - \log _2 \det \left( {{\bf{C}}_l  - {\bf{A}}_l \left( {{\bf{R}}_{g,\,l} \left( {{\bf{B}}_l  + \sum\limits_{t = l + 1}^L {{\bf{\Sigma }}_t } } \right) + N_0 {\bf{I}}_{N_t } } \right)^{ - 1} {\bf{R}}_{g,\,l} {\bf{A}}_l^H } \right)
 \end{array}
\end{equation}
\begin{equation}\label{achievable_rate_F3}
\begin{array}{l}
\qquad \quad \ \ = \log _2 \det \left( {{\bf{\Sigma }}_l } \right)  - \log _2 \det \left( \mathbf{N}_l \right). \qquad \qquad \qquad \qquad \qquad \qquad \qquad \qquad \qquad \qquad \qquad \
 \end{array}
\end{equation}
 \hrulefill
\vspace*{4pt}
\end{figure*}

\begin{equation}\label{achievable_rate_F4}
\begin{array}{l}
\mathbf{N}_l = {\bf{F}}_l \left( {{\bf{\Sigma }}_{S,\,l}  - {\bf{\Sigma }}_{S,\,l} {\bf{W}}_{ l} {\bf{\Sigma }}_{S,\,l} } \right){\bf{F}}_l^H  - {\bf{F}}_l {\bf{\Sigma }}_{S,\,l} {\bf{W}}_{l} {\bf{\Sigma }}_l  \\
\hspace{2cm}   - {\bf{\Sigma }}_l {\bf{W}}_{l} {\bf{\Sigma }}_{S,\,l} {\bf{F}}_l^H  + {\bf{\Sigma }}_l  - {\bf{\Sigma }}_l {\bf{W}}_{l} {\bf{\Sigma }}_l
\end{array}
\end{equation}
\begin{align}\label{achievable_rate_F5}
{\bf{W}}_{ l} & = \left( {{\bf{R}}_{g,\,l} \left( {{\bf{B}}_l  + \sum\limits_{t = l + 1}^L {{\bf{\Sigma }}_t } } \right) + N_0 {\bf{I}}_{N_t } } \right)^{ - 1} {\bf{R}}_{g,\,l} \\
& =  \left( {\left( {{\bf{B}}_l  + \sum\limits_{t = l + 1}^L {{\bf{\Sigma }}_t } } \right) + N_0 {\bf{R}}_{g,\, l}^{ - 1} } \right)^{ - 1}.
\end{align}

Now, we find ${\bf{F}}_l$ maximizing $ R_{{\rm{upp}},\, l}$.  Applying similar steps as those in (\ref{C_F_2})--(\ref{R_F_2}), we can obtain
\begin{equation}\label{achievable_rate_F6}
\begin{array}{l}
\left\{ {\nabla _{{\bf{F}}_l } R_{{\rm{upp}},\,l} } \right\}_{i,\, j}  \! =  \! - {\rm{tr}}\left( {{\bf{N}}_l^{ - 1} \left[ {{\bf{F}}_l \left( {{\bf{\Sigma }}_{S,\,l}  \! -  \! {\bf{\Sigma }}_{S,\,l} {\bf{W}}_l {\bf{\Sigma }}_{S,\,l} } \right){\bf{e}}_j {\bf{e}}_i^H } \right. } \right.\\
\hspace{5cm} \left. { \left. {- {\bf{\Sigma }}_l {\bf{W}}_l {\bf{\Sigma }}_{S,\,l} {\bf{e}}_j {\bf{e}}_i^H } \right]} \right).
\end{array}
\end{equation}

Then, we have a necessary condition for the optimal $\mathbf{F}_l$
\begin{equation}\label{achievable_rate_F7}
\begin{array}{l}
\nabla _{{\bf{F}}_l } R_{{\rm{upp}},\,l}  = - {\bf{N}}_l^{ - 1} \left[ {{\bf{F}}_l \left( {{\bf{\Sigma }}_{S,\,l}  - {\bf{\Sigma }}_{S,\,l} {\bf{W}}_l {\bf{\Sigma }}_{S,\,l} } \right) }\right. \\
\hspace{5cm}  \left. {- {\bf{\Sigma }}_l {\bf{W}}_l {\bf{\Sigma }}_{S,\,l} } \right] = \mathbf{0}.
\end{array}
\end{equation}

From (\ref{achievable_rate_F7}), a closed-form expression for the matrix $\mathbf{\widetilde{F}}_l$ can be derived as
\begin{align}\label{achievable_rate_F8}
{\bf{\widetilde{F} }}_l  & = {\bf{\Sigma }}_l {\bf{W}}_l \left( {{\bf{I}}_{N_t }  - {\bf{\Sigma }}_{S, \, l} {\bf{W}}_l } \right)^{ - 1} \\
\label{achievable_rate_F9}
&= {\bf{\Sigma }}_l \left( {{\bf{W}}_l^{ - 1}  - {\bf{\Sigma }}_{S,\,l} } \right)^{-1} \\
\label{achievable_rate_F10}
& = {\bf{\Sigma }}_l \left( {{\bf{\Sigma }}_l  + \sum\limits_{t = l + 1}^L {{\bf{\Sigma }}_t }  + N_0 {\bf{R}}_{g,\,l}^{ - 1} } \right)^{ - 1} .
\end{align}

Equation (\ref{achievable_rate_F10}) presents a unique closed-form structure of ${\bf{\widetilde{F} }}_l$ satisfying the necessary condition in (\ref{achievable_rate_F7}). Therefore,
the derived ${\widetilde{{\bf{F}}}}_l$ is a global maximizer of the upper bound  $R_{{\rm{upp}},\, l}$.

Next, by substituting the expression of $\mathbf{\widetilde{F}}_l$ in (\ref{achievable_rate_F10}) into $\mathbf{N}_l$ in (\ref{achievable_rate_F4}), it yields
\begin{align}\label{CPower_G1}
 {\bf{N}}_l & = \left( {{\bf{I}}_{N_t} - {\bf{\Sigma }}_l {\bf{W}}_l  - {{\bf{\widetilde{F}}}_{l}} {\bf{\Sigma }}_{S,\,l} {\bf{W}}_l } \right){\bf{\Sigma }}_l  \\
 \label{CPower_G2}
 & = \left( {{\bf{I}}_{N_t} - {\bf{\Sigma }}_l \left( {{\bf{I}}_{N_t} + \left( {{\bf{\Sigma}} _l  +  \sum\limits_{t = l + 1}^L {{\bf{\Sigma }}_t }  + N_0 {\bf{R}}_{g,\,l}^{ - 1} } \right)^{ - 1} }\right. } \right. \nonumber \\
& \hspace{5cm} \times \left. {\left. { {\bf{\Sigma }}_{S,\,l} } \right){\bf{W}}_l } \right){\bf{\Sigma }}_l  \\
  \label{CPower_G3}
&  = \left( {{\bf{I}}_{N_t}\! - \! {\bf{\Sigma}}_l \left( {{\bf{\Sigma}} _l \! + \! \sum\limits_{t = l + 1}^L {{\bf{\Sigma }}_t } \! + \! N_0 {\bf{R}}_{g,\,l}^{ - 1} } \right)^{ - 1} } {{\bf{W}}_l^{ - 1} {\bf{W}}_l }\right){\bf{\Sigma }}_l  \\
 \label{CPower_G4}
 & = \! \left( {{\bf{\Sigma}} _l \! + \! \sum\limits_{t = l + 1}^L {{\bf{\Sigma }}_t } \! + \! N_0 {\bf{R}}_{g,\,l}^{ - 1} } \right)^{ - 1}\!\! \left( {\sum\limits_{t = l + 1}^L {{\bf{\Sigma }}_t } \! + \! N_0 {\bf{R}}_{g,\,l}^{ - 1} } \right)\! {\bf{\Sigma }}_l .
\end{align}

Thus, the achievable rate of the $l$-th receiver can be upper bounded by
\begin{align}\label{CPower_rate_2}
R_l  \le \widetilde{R}_{{\rm{upp}},\,l} & =  \log _2 \det \left( { {\bf{\Sigma }}_l +  \sum\limits_{t = l + 1}^L {{\bf{\Sigma }}_t }  + N_0 {\bf{R}}_{g,\,l}^{-1} } \right) \nonumber \\
& \hspace{0.5cm} - \log _2 \det \left( { \sum\limits_{t = l + 1}^L {{\bf{\Sigma }}_t }  + N_0 {\bf{R}}_{g,\,l}^{-1} } \right) \\
 & = \log _2 \det \left( {{\bf{R}}_{g,\,l} \sum\limits_{t = l}^L {{\bf{\Sigma }}_t }  + N_0 {\bf{I}}_{N_t } } \right) \nonumber \\
& \hspace{0.5cm} - \log _2 \det \left( {{\bf{R}}_{g,\,l} \sum\limits_{t = l + 1}^L {{\bf{\Sigma }}_t }  + N_0 {\bf{I}}_{N_t } } \right).
\end{align}

\section{Proof of the Proposition  \ref{upper_bound_one}}\label{proof:upper_bound_one}
We note that when $N_t = 1$, the  power allocation design to maximize  $\widetilde R_{\rm upp,\,sum}^w$  lies either on the stationary points or the boundary points of the feasible solution set.  For the stationary points, we consider the KKT conditions. To this end, let $\nu$ be the Lagrange multiplier for the  constraint $\sum\nolimits_{t = 1}^L {P_t   } \leq P$.  Then,  necessary KKT conditions satisfied by the optimal solution can be written as
\begin{equation}\label{Power_KKT_1}
\frac{{\partial \widetilde R_{\rm upp,\,sum}^w  }}{{\partial P_t }} - \nu  = 0, \quad t = 1,2,\cdots,L
\end{equation}
where $\frac{{\partial \widetilde R_{\rm upp,\,sum}^w  } }{{\partial P_t }} = \sum\limits_{l = 1}^L {\frac{{\partial \widetilde{R}_{{\rm{upp}},\,l} }}{{\partial P_t }}}$ denotes the partial derivative of $\widetilde R_{\rm upp,\,sum}^w$ with respect to $P_t$, $t=1,2, \cdots, L$. The derivative of $\frac{{{\partial \widetilde{R}_{{\rm{upp}},\,l} } }}{{\partial P_t }}$ can be expressed as
\begin{equation}\label{Power_KKT_3}
\frac{{\partial { \widetilde{R}_{{\rm{upp}},\,l} } }}{{\partial P_t }} = \left\{ \begin{array}{lll}
 \frac{{r_l }}{{r_l P_l  + a_l }} - \frac{{r_l }}{{a_l }}, &{l < t}& \\
 \frac{{r_t }}{{r_t P_t  + a_t }}, &{l = t}& \\
 0, &{l > t}& \\
 \end{array} \right.
\end{equation}
where $a_l  = r_l\sum\nolimits_{t = l + 1}^L {P_t }  + N_0$. Substituting (\ref{Power_KKT_3}) into (\ref{Power_KKT_1}) yields
\begin{equation}\label{Power_KKT_4}
\frac{{r_t }}{{r_t P_t  + a_t }} + \sum\limits_{l = 1}^{t - 1} {\left( {\frac{{r_l }}{{r_l P_l  + a_l }} - \frac{{r_l }}{{a_l }}} \right)}  = \nu .
\end{equation}
For arbitrary $t = n$ and $t = n+1$ in (\ref{Power_KKT_4}), the following results hold
\begin{equation}\label{Power_KKT_5}
\nu = \frac{{r_t }}{{r_t P_t  + a_t }} + \sum\limits_{l = 1}^{t - 1} {\left( {\frac{{r_l }}{{r_l P_l  + a_l }} - \frac{{r_l }}{{a_l }}} \right)}
\end{equation}
and
\begin{equation}\label{Power_KKT_6}
\nu = \frac{{r_{t+1} }}{{r_{t+1} P_{t+1}  + a_{t+1} }} + \sum\limits_{l = 1}^{t } {\left( {\frac{{r_l }}{{r_l P_l  + a_l }} - \frac{{r_l }}{{a_l }}} \right)} .
\end{equation}
Subtracting (\ref{Power_KKT_5}) from (\ref{Power_KKT_6}), we have
\begin{equation}\label{Power_KKT_7}
\frac{{r_{t + 1} }}{{r_{t + 1} P_{t + 1}  + a_{t + 1} }} = \frac{{r_t }}{{a_t }} .
\end{equation}
Plugging the expressions of $a_t$ and  $a_{t+1}$ into (\ref{Power_KKT_7}) and simplifying yields
\begin{equation}\label{Power_KKT_8}
r_t = r_{t+1}, \quad t =1,2, \cdots, L-1 .
\end{equation}
The KKT conditions hold only when the equations in (\ref{Power_KKT_8}) are satisfied. When these equations are satisfied, the power allocation in (\ref{optimal_power}) will obviously maximize $\widetilde R_{\rm upp,\,sum}^w$. And when these equations are not satisfied, we note that the KKT conditions do not hold under any power allocation solutions. Hence the maximum  $\widetilde R_{\rm upp,\,sum}^w$  is achieved by boundary points (to allocate the total power to a certain receiver). Then, we know that selecting the receiver with the most robust channel condition achieves the maximum $\widetilde R_{\rm upp,\,sum}^w$, which completes the proof.


\begin{IEEEbiography}[{\includegraphics[width=1in,height=1.25in,clip,keepaspectratio]{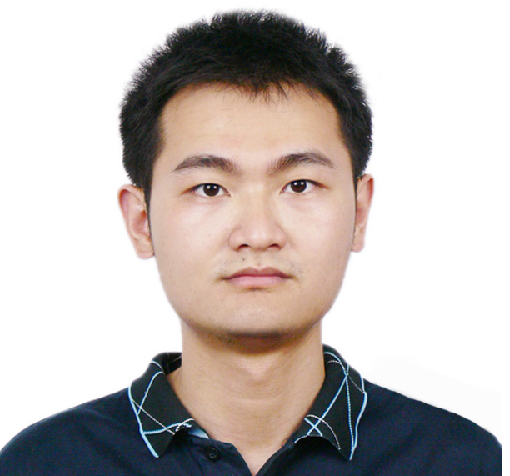}}]
{Yongpeng Wu} (S'08--M'13)
received the B.S. degree in telecommunication engineering from Wuhan University, Wuhan, China, in July 2007,
the Ph.D. degree in communication and signal processing
with the National Mobile Communications Research Laboratory,
Southeast University, Nanjing, China, in November 2013.

He is currently a Humboldt post-doc research fellow with Institute for Digital Communications, Universit$\ddot{a}$t Erlangen-N$\ddot{u}$rnberg,
Germany.  During his doctoral studies, he has conducted cooperative
research at the Department of Electrical Engineering, Missouri University of Science and Technology, USA. His
research interests include MIMO systems, signal processing for wireless communications, and
multivariate statistical theory.
\end{IEEEbiography}

\begin{IEEEbiography}[{\includegraphics[width=1in,height=1.25in,clip,keepaspectratio]{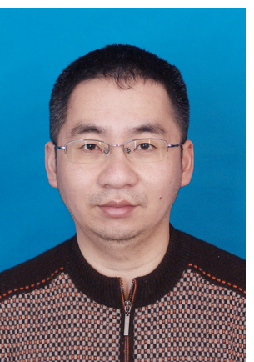}}]
{Shi Jin}
(S'06--M'07) received the B.S. degree in communications engineering from Guilin University
of Electronic Technology, Guilin, China, in 1996, the
M.S. degree from Nanjing University of Posts and
Telecommunications, Nanjing, China, in 2003, and
the Ph.D. degree in communications and information
systems from the Southeast University, Nanjing, in
2007. From June 2007 to October 2009, he was a
Research Fellow with the Adastral Park Research
Campus, University College London, London, U.K.
He is currently with the faculty of the National
Mobile Communications Research Laboratory, Southeast University. His
research interests include space time wireless communications, random matrix
theory, and information theory. He serves as an Associate Editor for the
IEEE Transactions on Wireless Communications, and IEEE Communications
Letters, and IET Communications. Dr. Jin and his co-authors have been
awarded the 2011 IEEE Communications Society Stephen O. Rice Prize Paper
Award in the field of communication theory and a 2010 Young Author Best
Paper Award by the IEEE Signal Processing Society.
\end{IEEEbiography}

\begin{IEEEbiography}[{\includegraphics[width=1in,height=1.25in,clip,keepaspectratio]{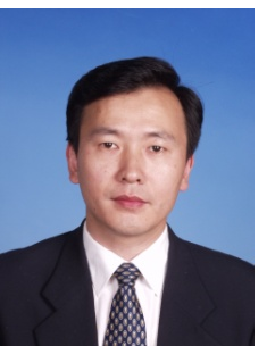}}]
{Xiqi Gao} received the Ph.D. degree in electrical engineering from Southeast University, Nanjing, China, in 1997.

He joined the Department of Radio Engineering, Southeast University, in April 1992. Since May 2001, he has been a professor of information systems and communications. From September 1999 to August 2000, he was a visiting scholar at Massachusetts Institute of Technology, Cambridge, and Boston University, Boston, MA. From August 2007 to July 2008, he visited the Darmstadt University of Technology, Darmstadt, Germany, as a Humboldt scholar. His current research interests include broadband multicarrier communications, MIMO wireless communications, channel estimation and turbo equalization, and multirate signal processing for wireless communications. From 2007 to 2012, he served as an Editor for the IEEE Transactions on Wireless Communications. From 2009 to 2013, he served as an Associate Editor for the IEEE Transactions on Signal Processing.

Dr. Gao received the Science and Technology Awards of the State Education Ministry of China in 1998, 2006 and 2009, the National Technological Invention Award of China in 2011, and the 2011 IEEE Communications Society Stephen O. Rice Prize Paper Award in the field of communications theory.

\end{IEEEbiography}

\begin{IEEEbiography}[{\includegraphics[width=1in,height=1.25in,clip,keepaspectratio]{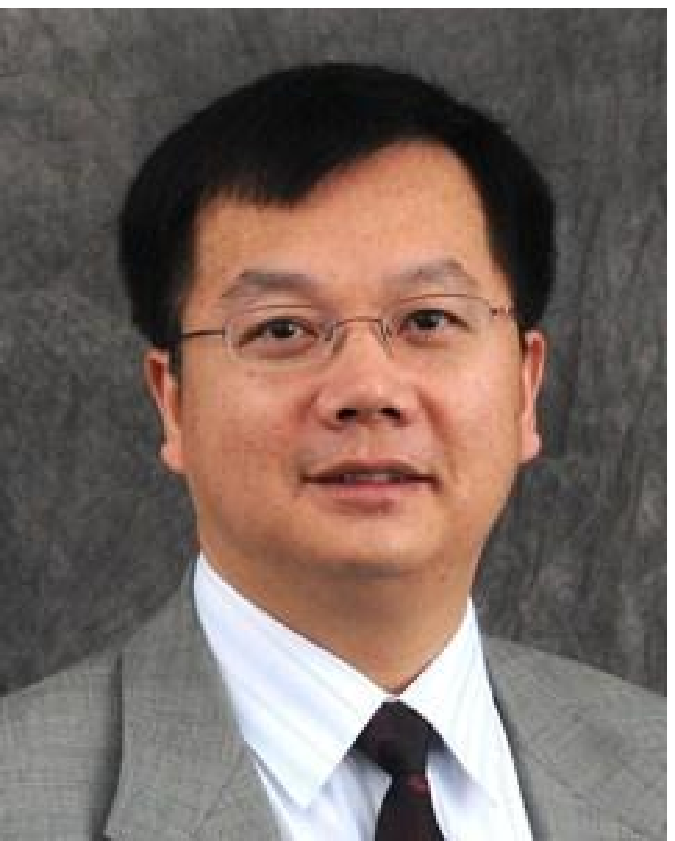}}]
{Chengshan Xiao} (M'99--SM'02--F'10)  received a bachelor of science degree in electronic engineering from the University of Electronic Science and Technology of China, Chengdu, China, in 1987, a master of science degree in electronic engineering from Tsinghua University, Beijing, China, in 1989, and a Ph.D. in electrical engineering from the University of Sydney, Sydney, Australia, in 1997.

From 1989 to 1993, he was a faculty member with the Department of Electronic Engineering, Tsinghua University. From 1997 to 1999, he was a Senior Member of Scientific Staff, Nortel Networks, Ottawa, Canada. From 1999 to 2000, he was a Faculty Member with the University of Alberta, Edmonton, AB, Canada. From 2000 to 2007, he was with the University of Missouri, Columbia, where he was an Assistant Professor and then an Associate Professor. He is currently a Professor with the Department of Electrical and Computer Engineering, Missouri University of Science and Technology, Rolla (formerly University of Missouri, Rolla). His current research interests include wireless communications, signal processing, and underwater acoustic communications. He is the holder of three U.S. patents. His invented algorithms have been implemented into Nortel's base station radios after successful technical field trials and network integration.

Dr. Xiao is the Director of Conference Publication of IEEE Communications Society (ComSoc), an Elected Member of IEEE ComSoc Board of Governors, a Member of IEEE ComSoc Fellow Evaluation Committee, and a Distinguished Lecturer of the IEEE Vehicular Technology Society. Previously, he served as an Editor, an Area Editor and the Editor-in-Chief of the IEEE Transactions on Wireless Communications; an Associate Editor of the IEEE Transactions on Vehicular Technology, the IEEE Transactions on Circuits and Systems-I, and the international journal of Multidimensional Systems and Signal Processing. He was the Technical Program Chair of the 2010 IEEE International Conference on Communications (ICC), the Lead Co-chair of the 2008 IEEE ICC Wireless Communications Symposium, and a Phy/MAC Program Co-chair of the 2007 IEEE Wireless Communications and Networking Conference. He served as the founding Chair of the IEEE Technical Committee on Wireless Communications and the Vice-Chair of the IEEE Technical Committee on Personal Communications. He is a recipient of the 2014 Humboldt Research Award.

\end{IEEEbiography}

\begin{IEEEbiography}[{\includegraphics[width=1in,height=1.25in,clip,keepaspectratio]{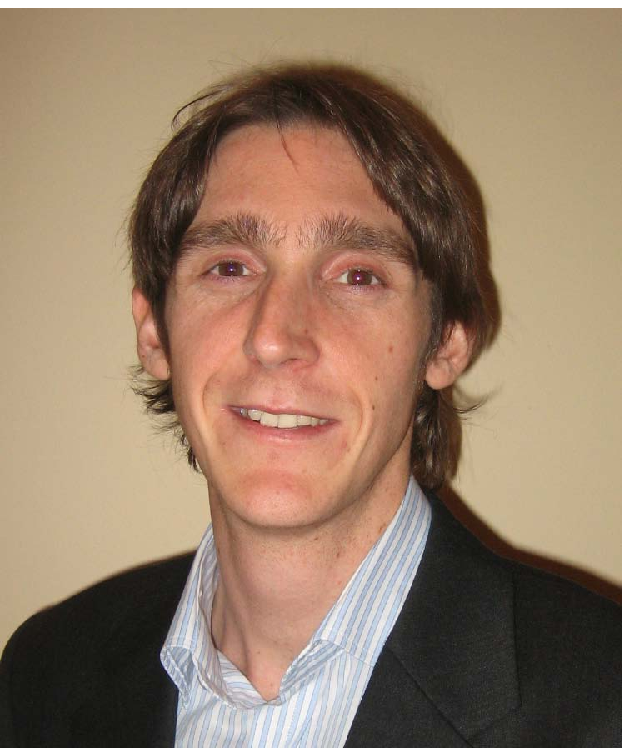}}]
{Matthew McKay} (S'03--M'07--SM'13) received a joint bachelor degree in Electrical Engineering and Computer Science from the Queensland University of Technology, Australia, in 2002, and the Ph.D. degree in Electrical Engineering from the University of Sydney, Australia, in 2007. Subsequently, he joined HKUST, where he is the Hari Harilela Associate Professor of Electronic and Computer Engineering. He is also affiliated with the Division of Biomedical Engineering. His research interests include broad topics in communications and signal processing. More recently, he has developed a keen interest in the interdisciplinary areas of computational immunology and financial engineering.

Matthew was awarded the University Medal upon graduating from the Queensland University of Technology. He along with his co-authors and students have received best paper awards at IEEE ICASSP 2006, IEEE VTC 2006, ACM IWCMC 2010, IEEE Globecom 2010, and IEEE ICC 2011. He also received a 2010 Young Author Best Paper Award by the IEEE Signal Processing Society, the 2011 Stephen O. Rice Prize in the Field of Communication Theory by the IEEE Communication Society, and the 2011 Young Investigator Research Excellence Award by the School of Engineering at HKUST. In 2013, he was the recipient the Asia-Pacific Best Young Researcher Award by the IEEE Communication Society.

\end{IEEEbiography}

\end{document}